\documentclass[english,superscriptaddress]{revtex4-1}
\usepackage[T1]{fontenc}
\usepackage[latin9]{inputenc}
\usepackage{color}
\usepackage{babel}
\usepackage{verbatim}
\usepackage{amsmath}
\usepackage{amssymb}
\usepackage{graphicx}
\usepackage[unicode=true,pdfusetitle,
 bookmarks=true,bookmarksnumbered=false,bookmarksopen=false,
 breaklinks=false,pdfborder={0 0 1},backref=false,colorlinks=true]
 {hyperref}

\makeatletter
\allowdisplaybreaks

\makeatother

\begin{document}
\title{Linear response theory for spin alignment of vector mesons in thermal
media}
\author{Wen-Bo Dong }
\affiliation{Department of Modern Physics, University of Science and Technology
of China, Hefei, Anhui 230026, China}
\author{Yi-Liang Yin}
\affiliation{Department of Modern Physics, University of Science and Technology
of China, Hefei, Anhui 230026, China}
\author{Xin-Li Sheng}
\affiliation{INFN Sezione di Firenze, Via Giovanni Sansone 1, 50019, Sesto Fiorentino
FI, Italy}
\author{Shi-Zheng Yang}
\affiliation{Department of Modern Physics, University of Science and Technology
of China, Hefei, Anhui 230026, China}
\author{Qun Wang}
\affiliation{Department of Modern Physics, University of Science and Technology
of China, Hefei, Anhui 230026, China}
\affiliation{School of Mechanics and Physics, Anhui University of Science and Technology,
Huainan,Anhui 232001, China}
\begin{abstract}
We present a calculation of the spin alignment for unflavored vector
mesons in thermalized quark-gluon plasma based on the Kubo formula
in linear response theory. This is achieved by expanding the system
to the first order of the coupling constant and the spatial gradient.
The effect strongly relies on the vector meson's spectral functions
which are determined by the interaction and medium properties. The
spectral functions are calculated for the one-quark-loop self-energy
with meson-quark interaction. The numerical results show that the
correction to the spin alignment from the thermal shear tensor is
of the order {\normalsize{}$10^{-4}\sim10^{-5}$} for the chosen values
of quark-meson coupling constant, if the magnitude of thermal shear
tensor is {\normalsize{}$10^{-2}$}.
\end{abstract}
\maketitle
% done, QW, 2023.11.24, 3:40

\section{Introduction}

Rotation and spin polarization are inherently connected and can be
converted to each other as demonstrated in the Barnett effect \citep{Barnett:1935}
and Einstein-de Haas effect \citep{dehaas:1915} in materials. The
same phenomenon known as global polarization can also exist in peripheral
heavy-ion collisions at high energies in which the huge orbital angular
momentum is partially distributed into the strong interaction matter
in the form of particles' spin polarization \citep{Liang:2004ph,Voloshin:2004ha,Betz:2007kg,Becattini:2007sr,Gao:2007bc}.
The global polarization of hyperons has been observed in experiments
\citep{STAR:2017ckg,STAR:2018gyt} and been extensively studied in
recent years \citep{Wang:2017jpl,Florkowski:2018fap,Becattini:2020ngo,Gao:2020lxh,Huang:2020dtn,Gao:2023chinphyb}.

% done, QW, 2023.10.20, 5:00

Unlike the spin polarization of hyperons that can be measured through
their weak decay, vector mesons can only decay by strong interaction
which respects parity symmetry, which makes their spin polarization
inaccessible in experiments. For spin-1 vector mesons, the only spin
observables that can be measured are some elements of the spin density
matrix $\rho_{\lambda_{1}\lambda_{2}}$ with $\lambda_{1}$ and $\lambda_{2}$
denoting spin states along the spin quantization direction. One of
them is $\rho_{00}$ that can be measured through the decay daughter's
polar angle distribution in the rest frame of the vector meson. If
$\rho_{00}$ is not 1/3, it means that the spin-0 state is not equally
occupied among three spin states, which is called the spin alignment.
The global spin alignment in heavy-ion collisions was first suggested
by Liang and Wang \citep{Liang:2004xn}. The global spin alignments
of $\phi$ and $K^{*0}$ mesons were first measured by STAR collaboration
in Au+Au collisions $\sqrt{s_{NN}}=200$ GeV in 2008 \citep{STAR:2008lcm},
but no signals were found. With the accumulation of experimental data,
STAR Collaboration finally found a large spin alignment for $\phi$
mesons in Au+Au collision at lower energies but not for $K^{*0}$
\citep{STAR:2022fan}.

% done, QW, 2023.10.20, 8:50

Such a large spin alignment for $\phi$ mesons cannot be fully accounted
by conventional mechanism \citep{Yang:2017sdk,Xia:2020tyd,Gao:2021rom,Muller:2021hpe,Kumar:2023ghs}.
Some of us proposed that local fluctuations of vector fields in strong
interaction may give a large deviation of $\rho_{00}$ from 1/3 for
$\phi$ mesons \citep{Sheng:2019kmk}. Such a prediction was made
in a nonrelativistic quark coalescence model \citep{Yang:2017sdk,Sheng:2020ghv}
that works for static or nearly static mesons in principle. Such a
nonrelativistic quark coalescence model has been promoted to a relativistic
version \citep{Sheng:2022wsy,Sheng:2022ffb} based on quantum transport
theory \citep{Chou:1984es,Blaizot:2001nr,Berges:2004yj,Cassing:2008nn}
with the help of covariant Wigner functions for massive particles
\citep{Heinz:1983nx,Vasak:1987um,Zhuang:1995pd,Fang:2016vpj,Gao:2019znl,Weickgenannt:2019dks,Weickgenannt:2020aaf,Weickgenannt:2021cuo}
and matrix valued spin-dependent distributions \citep{Becattini:2013fla,Sheng:2021kfc}.
With fluctuation parameters of strong interaction fields extracted
from transverse momentum-integrated data for $\rho_{00}$ as a function
of the collision energy, the calculated transverse momentum dependence
of $\rho_{00}$ agrees with STAR's data for $\phi$ mesons \citep{STAR:2022fan}.
The rapidity dependence of $\rho_{00}$ has also been predicted with
same parameters before preliminary data of STAR was released: the
main feature of the data can be described by the theoretical result
\citep{Sheng:2023urn}. For recent reviews on the spin alignment of
vector mesons, see, e.g., Refs. \citep{Chen:2023hnb,Wang:2023fvy,Sheng:2023chinphyb}.

% done, QW, 2023.10.20, 12:00

Recently, the contribution from the thermal shear tensor to the spin
alignment of the vector meson has been calculated using the linear
response theory \citep{Li:2022vmb} and kinetic theory \citep{Wagner:2022gza}.
The authors of Ref. \citep{Li:2022vmb} argued that this contribution
is quite large based on an estimate of the energy shift and width
of the vector meson in medium without really calculating them. This
work was inspired by Refs. \citep{Yi:2021ryh,Becattini:2021suc,Fu:2021pok,Niida:2018hfw,Becattini:2021iol}
pointing out that there is a coupling between the spin polarization
and the thermal shear tensor which can partially resolve the local
polarization puzzle of $\Lambda$ hyperons.

% done, QW, 2023.10.20, 17:00

In this paper, we will calculate the spin alignment of vector mesons
from the Kubo formula in linear response theory \citep{zubarev1996statistical,zubarev1997statistical,Zubarev_1979,Kapusta:2006pm}
in thermalized quark-gluon plasma (QGP). Vector mesons are assumed
to be thermalized, and quarks and antiquarks are assumed to be unpolarized.
The interaction is described by the vertex between vector meson and
quark-antiquark \citep{Manohar:1983md,Fernandez:1993hx,Zacchi:2015lwa,Zacchi:2016tjw}.
In Sec. \ref{sec:green-functions}, we present two-point Green's functions
of different kinds for vector mesons in the closed-time-path (CTP)
formalism \citep{Chou:1984es,Blaizot:2001nr,Wang:2001dm,Berges:2004yj,Cassing:2008nn,Crossley:2015evo}.
In Sec. \ref{sec:wigner-functions}, we give an introduce on spin
density matrices for vector mesons from Wigner functions. In Sec.
\ref{sec:ds-equation}, we present the Dyson-Schwinger equation for
retarded Green's functions. We give the expression for retarded self-energies
of vector mesons including the contribution from one-quark-loop. In
Sec. \ref{sec:Self-Energy-and-Spectral}, we give the general form
of spectral functions in medium for vector mesons from retarded Green's
functions. In Sec. \ref{sec:Kubo-Formula}, we use the Kubo formula
in the linear response theory \citep{zubarev1996statistical,zubarev1997statistical,Zubarev_1979,Kapusta:2006pm}
to calculate the correction to the two-point Green's function proportional
to the thermal shear tensor. From it we are able to calculate the
correction to $\rho_{00}$ in Sec. \ref{sec:Self-Energy-and-Shear}.
We adopt the hard-thermal-loop (HTL) \citep{Weldon:1982aq,Pisarski:1988vd,Braaten:1989mz,Thoma:2000dc,Blaizot:2001nr,Bellac:2011kqa}
and quasi-particle approximations \citep{Gale:1990pn} to calculate
spectral functions. The HTL approximation provides a toy model to
illustrate the physics inside this problem since we have analytical
formula for spectra functions. Then we consider a more realistic quasi-particle
approximation for spectral functions. Under a few approximations or
assumptions, we obtain an analytical expression for the correction
to $\rho_{00}$, which depends on the width and energy shift from
the self-energy. The numerical results for the tensor coefficients
in the correction to $\rho_{00}$ are presented. The conclusion and
discussion are given in Sec. \ref{sec:Conclusion-and-Discussion}.

% done, QW, 2023.10.23, 9:00

In this paper, we adopt following notational conventions: $g^{\mu\nu}=\mathrm{diag}(1,-1,-1,-1)$
where $\mu,\nu=0,1,2,3$, $x^{\mu}=(x^{0},\mathbf{x})=(x_{0},\mathbf{x})$,
$x\cdot y=x^{\mu}y_{\mu}$, $x_{(\mu}y_{\nu)}=(1/2)(x_{\mu}y_{\nu}+x_{\nu}y_{\mu})$,
$\hbar=k_{B}=1$. Greek letters denote components of four-vectors
while lowercase Latin letters as subscripts denote components three-vectors.
The four-momentum $p^{\mu}$ is not necessarily on-shell unless we
add an index 'on'. The summation of repeated indices is implied if
not stated explicitly. The definition of two-point Green's functions
$G$ and $\Sigma$ in this paper differs by a factor $i=\sqrt{-1}$
from the usual one in quantum field theory, which are related by $G=i\widetilde{G}$
and $\Sigma=i\widetilde{\Sigma}$.

% done, QW, 2023.10.24, 8:30

\section{Two-point Green's functions}

\label{sec:green-functions}
\begin{figure}
\includegraphics[scale=0.6]{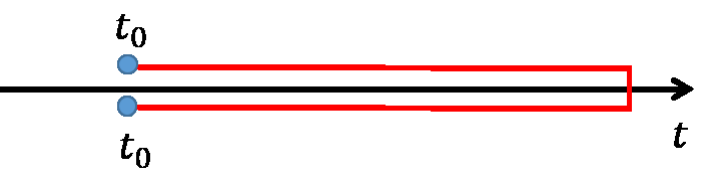}

\caption{Illustration of the closed-time-path upon which the non-equilibrium
quantum field theory is built. \label{fig:ctp}}
\end{figure}

In this section we will give an introduction
to two-point Green's functions for vector mesons on the CTP as shown
in Fig. \ref{fig:ctp}. The CTP formalism is a field-theory based
method for many-body systems in off-equilibrium as well in equilibrium
\citep{Chou:1984es,Blaizot:2001nr,Wang:2001dm,Berges:2004yj,Cassing:2008nn,Crossley:2015evo}.
When it is used for systems in equilibrium, it is actually the real
time formalism of the thermal (finite temperature and density) field
theory \citep{Kapusta:2006pm,Kapusta:2023eix}. Wigner functions can
be obtained from two-point Green's functions and are related to spin
density matrices, which will be addressed in the next section. We
refer the readers to Section 2.2 of Ref. \citep{Hidaka:2022dmn} for
a very brief introduction to two-point Green's functions on the CTP.

% done, QW, 2024.2.14, 7:30

The Lagrangian density for unflavored vector mesons with spin-1 and
mass $m_{V}$ reads 
\begin{eqnarray}
\mathcal{L} & = & -\frac{1}{4}F_{\mu\nu}F^{\mu\nu}+\frac{m_{V}^{2}}{2}A_{\mu}A^{\mu}-A_{\mu}j^{\mu}.\label{eq:Lagrangian density}
\end{eqnarray}
where $A^{\mu}(x)$ is the real vector field for the meson, $F_{\mu\nu}=\partial_{\mu}A_{\nu}-\partial_{\nu}A_{\mu}$
is the field strength tensor, and $j^{\mu}$ is the source coupled
to $A^{\mu}(x)$.

% done, QW, 2024.2.15, 7:00

The two-point Green's function on the CTP is defined as
\begin{eqnarray}
G_{\mathrm{CTP}}^{\mu\nu}(x_{1},x_{2}) & = & \left\langle T_{C}A^{\mu}(x_{1})A^{\nu\dagger}(x_{2})\right\rangle ,
\end{eqnarray}
where $\left\langle \cdots\right\rangle $ denotes the ensemble average
and $T_{C}$ denotes time order operator on the CTP contour. Depending
on whether the field $A^{\mu}$ lives on the positive or negative
time branch, we have four components $G_{\mathrm{CTP}}^{\mu\nu}$,
\begin{eqnarray}
G_{F}^{\mu\nu}(x_{1},x_{2}) & \equiv & G_{++}^{\mu\nu}(x_{1},x_{2})\nonumber \\
 & = & \theta(t_{1}-t_{2})\left\langle A^{\mu}(x_{1})A^{\nu}(x_{2})\right\rangle +\theta(t_{2}-t_{1})\left\langle A^{\nu}(x_{2})A^{\mu}(x_{1})\right\rangle ,\nonumber \\
G_{<}^{\mu\nu}(x_{1},x_{2}) & = & G_{+-}^{\mu\nu}(x_{1},x_{2})=\left\langle A^{\nu}(x_{2})A^{\mu}(x_{1})\right\rangle ,\nonumber \\
G_{>}^{\mu\nu}(x_{1},x_{2}) & = & G_{-+}^{\mu\nu}(x_{1},x_{2})=\left\langle A^{\mu}(x_{1})A^{\nu}(x_{2})\right\rangle ,\nonumber \\
G_{\overline{F}}^{\mu\nu}(x_{1},x_{2}) & \equiv & G_{--}^{\mu\nu}(x_{1},x_{2})\nonumber \\
 & = & \theta(t_{2}-t_{1})\left\langle A^{\mu}(x_{1})A^{\nu}(x_{2})\right\rangle +\theta(t_{1}-t_{2})\left\langle A^{\nu}(x_{2})A^{\mu}(x_{1})\right\rangle .\label{eq:def_4}
\end{eqnarray}
From the constraint $G_{F}^{\mu\nu}+G_{\overline{F}}^{\mu\nu}=G_{<}^{\mu\nu}+G_{>}^{\mu\nu}$,
only three of them are independent. In the so-called physical representation
\citep{Chou:1984es,kadanoff1962quantum,fetter2003quantum}, three
independent two-point Green's functions are 
\begin{eqnarray}
G_{R}^{\mu\nu}(x_{1},x_{2}) & = & (G_{F}^{\mu\nu}-G_{<}^{\mu\nu})(x_{1},x_{2})\approx\theta(t_{1}-t_{2})(G_{>}^{\mu\nu}-G_{<}^{\mu\nu})(x_{1},x_{2}),\nonumber \\
G_{A}^{\mu\nu}(x_{1},x_{2}) & = & (G_{F}^{\mu\nu}-G_{>}^{\mu\nu})(x_{1},x_{2})\approx\theta(t_{2}-t_{1})(G_{<}^{\mu\nu}-G_{>}^{\mu\nu})(x_{1},x_{2}),\nonumber \\
G_{C}^{\mu\nu}(x_{1},x_{2}) & = & G_{>}^{\mu\nu}(x_{1},x_{2})+G_{<}^{\mu\nu}(x_{1},x_{2}),\label{eq:def_7}
\end{eqnarray}
where the subscripts $"A"$ and $"R"$ denote the advanced and retarded
Green's function respectively. The two-point Green's functions in
Eqs. (\ref{eq:def_4}-\ref{eq:def_7}) can be used to express any
two-point functions defined on the CTP contour such as the self energy
$\Sigma^{\mu\nu}(x_{1},x_{2})$. When dealing with the vacuum contributions
to $G_{R,A}^{\mu\nu}$, the last equalities in the first and second
line of Eq. (\ref{eq:def_7}) do not exactly hold since a singular
term $\sim\delta(t_{1}-t_{2})$ is missing.

% done, QW, 2023.10.2, 11:45

\section{Wigner functions and spin density matrices}

\label{sec:wigner-functions}In this section, we will introduce how
one can obtain spin density matrices for vector mesons from Wigner
functions. We refer the readers to some recent reviews \citep{Sheng:2023chinphyb,Becattini:2024uha}
for details of the topic.

The second quantization of the vector field is in the form 
\begin{eqnarray}
A^{\mu}(x) & = & \sum_{\lambda=0,\pm1}\int\frac{d^{3}p}{(2\pi)^{3}}\frac{1}{2E_{p}^{V}}\nonumber \\
 &  & \times\left[\epsilon^{\mu}(\lambda,{\bf p})a(\lambda,{\bf p})e^{-ip_{\mathrm{on}}\cdot x}+\epsilon^{\mu\ast}(\lambda,{\bf p})a^{\dagger}(\lambda,{\bf p})e^{ip_{\mathrm{on}}\cdot x}\right],\label{eq:a-quantization}
\end{eqnarray}
where $p_{\mathrm{on}}^{\mu}=(E_{p}^{V},\mathbf{p})$ is the on-shell
momentum of the vector meson, $E_{p}^{V}=\sqrt{|{\bf p}|^{2}+m_{V}^{2}}$
is the vector meson's energy, $\lambda$ denotes the spin state, $a(\lambda,{\bf p})$
and $a^{\dagger}(\lambda,{\bf p})$ are annihilation and creation
operators respectively, and $\epsilon^{\mu}(\lambda,{\bf p})\equiv\epsilon_{\mu}(\lambda,p_{\mathrm{on}})$
represents the polarization vector obeying the following relations
\begin{eqnarray}
p_{\mathrm{on}}^{\mu}\epsilon_{\mu}(\lambda,p_{\mathrm{on}}) & = & 0,\nonumber \\
\epsilon(\lambda,p_{\mathrm{on}})\cdot\epsilon^{*}(\lambda^{\prime},p_{\mathrm{on}}) & = & -\delta_{\lambda\lambda^{\prime}},\nonumber \\
\Sigma_{\lambda}\epsilon^{\mu}(\lambda,p_{\mathrm{on}})\epsilon^{\nu,*}(\lambda,p_{\mathrm{on}}) & = & -\Delta^{\mu\nu}(p_{\mathrm{on}}),
\end{eqnarray}
where $\Delta^{\mu\nu}(p)=g^{\mu\nu}-p^{\mu}p^{\nu}/p^{2}$ is the
projector perpendicular to $p^{\mu}$. One can check that the quantum
field $A^{\mu}$ defined in Eq. (\ref{eq:a-quantization}) is Hermitian,
$A^{\mu}=A^{\mu\dagger}$. 

The Wigner function can be defined from $G_{\mu\nu}^{<}(x_{1},x_{2})$
{[}or equivalently $G_{\mu\nu}^{>}(x_{1},x_{2})${]} by taking a Fourier
transform with respect to the relative position $y=x_{1}-x_{2}$,
\begin{eqnarray}
G_{\mu\nu}^{<}(x,p) & \equiv & \int d^{4}y\,e^{ip\cdot y}G_{\mu\nu}^{<}(x_{1},x_{2})\nonumber \\
 & = & \int d^{4}y\,e^{ip\cdot y}\left\langle A_{\nu}^{\dagger}(x_{2})A_{\mu}(x_{1})\right\rangle .\label{eq:definition_G<}
\end{eqnarray}
Inserting the quantized field (\ref{eq:a-quantization}) into the
definition of the Wigner function (\ref{eq:definition_G<}), we obtain
\begin{eqnarray}
G_{\mu\nu}^{(0)<}(x,p) & = & 2\pi\sum_{\lambda_{1},\lambda_{2}}\delta\left(p^{2}-m_{V}^{2}\right)\left\{ \theta(p^{0})\epsilon_{\mu}\left(\lambda_{1},{\bf p}\right)\epsilon_{\nu}^{\ast}\left(\lambda_{2},{\bf p}\right)f_{\lambda_{1}\lambda_{2}}^{(0)}(x,{\bf p})\right.\nonumber \\
 &  & \left.+\theta(-p^{0})\epsilon_{\mu}^{\ast}\left(\lambda_{1},-{\bf p}\right)\epsilon_{\nu}\left(\lambda_{2},-{\bf p}\right)\left[\delta_{\lambda_{2}\lambda_{1}}+f_{\lambda_{2}\lambda_{1}}^{(0)}(x,-{\bf p})\right]\right\} ,\label{eq:wigner-func-vm}
\end{eqnarray}
where the superscript ``(0)'' denotes the leading order contribution
in $\hbar$ or gradient expansion, and the MVSD \citep{Becattini:2013fla,Sheng:2021kfc}
at the leading order for the vector meson is defined as 
\begin{equation}
f_{\lambda_{1}\lambda_{2}}^{(0)}(x,{\bf p})\equiv\int\frac{d^{4}u}{2(2\pi)^{3}}\delta(p\cdot u)e^{-iu\cdot x}\left\langle a_{V}^{\dagger}\left(\lambda_{2},{\bf p}-\frac{{\bf u}}{2}\right)a_{V}\left(\lambda_{1},{\bf p}+\frac{{\bf u}}{2}\right)\right\rangle .
\end{equation}
Note that $f_{\lambda_{1}\lambda_{2}}^{(0)}(x,{\bf p})$ is actually
the (unnormalized) spin density matrix $\rho_{\lambda_{1}\lambda_{2}}$,
which can be decomposed into the scalar, polarization ($P_{i}$) and
tensor polarization ($T_{ij}$) parts as \citep{Sheng:2023chinphyb,Becattini:2024uha}
\begin{equation}
f_{\lambda_{1}\lambda_{2}}^{(0)}=\mathrm{Tr}(f^{(0)})\left(\frac{1}{3}+\frac{1}{2}P_{i}\Sigma_{i}+T_{ij}\Sigma_{ij}\right)_{\lambda_{1}\lambda_{2}},\label{eq:mvsd-decomp}
\end{equation}
where $i,j=1,2,3$, $\mathrm{Tr}(f^{(0)})=\sum_{\lambda}f_{\lambda\lambda}^{(0)}$,
and $\Sigma_{i}$ and $\Sigma_{ij}$ are $3\times3$ traceless matrices
defined as 
\begin{align}
\Sigma_{1}= & \frac{1}{\sqrt{2}}\left(\begin{array}{ccc}
0 & 1 & 0\\
1 & 0 & 1\\
0 & 1 & 0
\end{array}\right),\;\;\Sigma_{2}=\frac{1}{\sqrt{2}}\left(\begin{array}{ccc}
0 & -i & 0\\
i & 0 & -i\\
0 & i & 0
\end{array}\right),\;\;\Sigma_{3}=\left(\begin{array}{ccc}
1 & 0 & 0\\
0 & 0 & 0\\
0 & 0 & -1
\end{array}\right),\nonumber \\
\Sigma_{ij}= & \frac{1}{2}(\Sigma_{i}\Sigma_{j}+\Sigma_{j}\Sigma_{i})-\frac{2}{3}\delta_{ij}.\label{eq:sigma-matrix}
\end{align}

Let us define an integrated or on-shell Wigner function 
\begin{equation}
W^{\mu\nu}(x,p_{\mathrm{on}})=\frac{E_{p}}{\pi}\int_{0}^{\infty}dp_{0}G_{<}^{\mu\nu}(x,p)=\sum_{\lambda_{1},\lambda_{2}}\epsilon^{\mu}\left(\lambda_{1},{\bf p}\right)\epsilon^{\nu\ast}\left(\lambda_{2},{\bf p}\right)f_{\lambda_{1}\lambda_{2}}(x,{\bf p}).\label{eq:wigner-decomp}
\end{equation}
It is easy to check that the second equality holds for the leading
order Wigner function $G_{\mu\nu}^{(0)<}(x,p)$ given by Eq. (\ref{eq:wigner-func-vm}).
But we assume that it hold at any order. One can check that $W^{\mu\nu}(x,p_{\mathrm{on}})$
is always transverse to the on-shell momentum, $p_{\mu}^{\mathrm{on}}W^{\mu\nu}(x,p_{\mathrm{on}})=0$.
The on-shell Wigner function can be decomposed into the scalar ($\mathcal{S}$),
polorization ($W^{[\mu\nu]}$) and tensor polarization ($\mathcal{T}^{\mu\nu}$)
parts as \citep{Sheng:2023chinphyb,Becattini:2024uha}
\begin{equation}
W^{\mu\nu}(x,p_{\mathrm{on}})=W^{[\mu\nu]}+W^{(\mu\nu)}=-\frac{1}{3}\Delta^{\mu\nu}(p_{\mathrm{on}})\mathcal{S}+W^{[\mu\nu]}+\mathcal{T}^{\mu\nu},\label{eq:wigner-decomp-1}
\end{equation}
where each part is defined as 
\begin{align}
W^{[\mu\nu]}\equiv & \frac{1}{2}(W^{\mu\nu}-W^{\nu\mu}),\nonumber \\
W^{(\mu\nu)}\equiv & \frac{1}{2}(W^{\mu\nu}+W^{\nu\mu}),\nonumber \\
\mathcal{T}^{\mu\nu}\equiv & W^{(\mu\nu)}+\frac{1}{3}\Delta^{\mu\nu}(p_{\mathrm{on}})\mathcal{S}.
\end{align}
With Eq. (\ref{eq:wigner-decomp-1}) one can show that both $W^{[\mu\nu]}$
and $\mathcal{T}^{\mu\nu}$ are traceless, $g_{\mu\nu}W^{[\mu\nu]}=g_{\mu\nu}\mathcal{T}^{\mu\nu}=0$.
Inserting Eq. (\ref{eq:mvsd-decomp}) into Eq. (\ref{eq:wigner-decomp}),
we have 
\begin{align}
\mathcal{S}= & \mathrm{Tr}(f)=-\Delta^{\mu\nu}(p_{\mathrm{on}})W_{\mu\nu},\nonumber \\
W^{[\mu\nu]}= & \frac{1}{2}\mathrm{Tr}(f)\sum_{\lambda_{1},\lambda_{2}}\epsilon^{\mu}\left(\lambda_{1},{\bf p}\right)\epsilon^{\nu\ast}\left(\lambda_{2},{\bf p}\right)P_{i}\Sigma_{\lambda_{1}\lambda_{2}}^{i},\nonumber \\
\mathcal{T}^{\mu\nu}= & \mathrm{Tr}(f)\sum_{\lambda_{1},\lambda_{2}}\epsilon^{\mu}\left(\lambda_{1},{\bf p}\right)\epsilon^{\nu\ast}\left(\lambda_{2},{\bf p}\right)T_{ij}\Sigma_{\lambda_{1}\lambda_{2}}^{ij}.\label{eq:swt-mvsd}
\end{align}
We see that $W^{[\mu\nu]}$ is related to $P_{i}$ while $\mathcal{T}^{\mu\nu}$
is related to $T_{ij}$. 

We can extract $f_{00}\propto\rho_{00}$ by projecting 
\begin{equation}
L^{\mu\nu}(p_{\mathrm{on}})=\epsilon^{\mu,*}\left(0,{\bf p}\right)\epsilon^{\nu}\left(0,{\bf p}\right)+\frac{1}{3}\Delta^{\mu\nu}(p_{\mathrm{on}}),\label{eq:l-munu-on}
\end{equation}
onto $W^{\mu\nu}$ in Eq. (\ref{eq:wigner-decomp}) as 
\begin{align}
L_{\mu\nu}(p_{\mathrm{on}})W^{\mu\nu}= & \sum_{\lambda_{1},\lambda_{2}}L_{\mu\nu}(p_{\mathrm{on}})\epsilon^{\mu}\left(\lambda_{1},{\bf p}\right)\epsilon^{\nu\ast}\left(\lambda_{2},{\bf p}\right)f_{\lambda_{1}\lambda_{2}}(x,{\bf p})\nonumber \\
= & f_{00}(x,{\bf p})+\frac{1}{3}\sum_{\lambda_{1},\lambda_{2}}\epsilon^{\mu}\left(\lambda_{1},{\bf p}\right)\epsilon_{\mu}^{\ast}\left(\lambda_{2},{\bf p}\right)f_{\lambda_{1}\lambda_{2}}(x,{\bf p})\nonumber \\
= & f_{00}(x,{\bf p})-\frac{1}{3}\mathrm{Tr}(f).\label{eq:l-munu}
\end{align}
In (\ref{eq:l-munu-on}), $\epsilon^{\mu}(0,{\bf p})$ is the polarization
vector along the spin quantization direction. With the first line
of Eq. (\ref{eq:swt-mvsd}) and Eq. (\ref{eq:l-munu}), we obtain
\begin{equation}
\frac{L_{\mu\nu}(p_{\mathrm{on}})W^{\mu\nu}}{-\Delta^{\mu\nu}(p_{\mathrm{on}})W_{\mu\nu}}=\frac{f_{00}(x,{\bf p})}{\mathrm{Tr}[f(x,{\bf p})]}-\frac{1}{3}=\rho_{00}-\frac{1}{3}.\label{eq:rho-00-1/3}
\end{equation}
The above formula relates the Wigner function to $\rho_{00}$, which
we will use to calculate the correction to $\rho_{00}$ in Sect. \ref{sec:Self-Energy-and-Shear}.

% done, QW and XLS, 2024.2.16, 11:00

\section{Dyson-Schwinger equation on CTP}

\label{sec:ds-equation}In this section we will give an introduction
to the Dyson-Schwinger equation (DSE) on the CTP which incorporates
retarded and advanced self-energies to be used for spectral functions
in the next section.

% done, QW, 2024.2.14, 8:30

We start from the integral form of the Dyson-Schwinger equation (DSE)
on the CTP for the vector meson \citep{Sheng:2022ffb,Wagner:2023cct}
\begin{eqnarray}
G^{\mu\nu}(x_{1},x_{2}) & = & G_{(0)}^{\mu\nu}(x_{1},x_{2})+\int_{C}dx_{1}^{\prime}dx_{2}^{\prime}G_{(0),\rho}^{\mu}(x_{1},x_{1}^{\prime})\Sigma_{\;\sigma}^{\rho}(x_{1}^{\prime},x_{2}^{\prime})G^{\sigma\nu}(x_{2}^{\prime},x_{2}),\label{eq:classical DSE}
\end{eqnarray}
where $dx_{1,2}^{\prime}\equiv d^{4}x_{1,2}^{\prime}$, $\int_{C}$
denotes the integral on the CTP contour, $G_{(0)}^{\mu\nu}$ and $G^{\mu\nu}$
are the bare and full propagator respectively, and $\Sigma^{\rho\sigma}$
is the self-energy. In Eq. (\ref{eq:classical DSE}) we have suppressed
the index 'CTP' in two-point functions $G_{(0)}^{\mu\nu}$, $G^{\mu\nu}$
and $\Sigma^{\rho\sigma}$. Contracting $(G_{(0)}^{\mu\nu})^{-1}$
on both sides of Eq. (\ref{eq:classical DSE}) and writing the DSE
in the matrix form, we obtain 
\begin{align}
 & -i\left[g_{\;\rho}^{\mu}(\partial_{x_{1}}^{2}+m_{V}^{2})-\partial_{x_{1}}^{\mu}\partial_{\rho}^{x_{1}}\right]\left(\begin{array}{cc}
G_{F}^{\rho\nu} & G_{<}^{\rho\nu}\\
G_{>}^{\rho\nu} & G_{\overline{F}}^{\rho\nu}
\end{array}\right)(x_{1},x_{2})\nonumber \\
= & \left(\begin{array}{cc}
1 & 0\\
0 & -1
\end{array}\right)g^{\mu\nu}\delta^{(4)}(x_{1}-x_{2})\nonumber \\
 & +\int dx^{\prime}\left(\begin{array}{cc}
\Sigma_{F,\rho}^{\mu} & -\Sigma_{<,\rho}^{\mu}\\
\Sigma_{>,\rho}^{\mu} & -\Sigma_{\overline{F},\rho}^{\mu}
\end{array}\right)(x_{1},x^{\prime})\left(\begin{array}{cc}
G_{F}^{\rho\nu} & G_{<}^{\rho\nu}\\
G_{>}^{\rho\nu} & G_{\overline{F}}^{\rho\nu}
\end{array}\right)(x^{\prime},x_{2}),\label{eq:DSE}
\end{align}
where the integral over $x_{2}^{\prime}$ is an normal one (not on
the CTP). Under a unitary transformation, Eq.(\ref{eq:DSE}) can be
put into the physical representation 
\begin{alignat}{1}
 & -i\left[g_{\;\rho}^{\mu}(\partial_{x_{1}}^{2}+m_{V}^{2})-\partial_{x_{1}}^{\mu}\partial_{\rho}^{x_{1}}\right]\left(\begin{array}{cc}
0 & G_{A}^{\rho\nu}\\
G_{R}^{\rho\nu} & G_{C}^{\rho\nu}
\end{array}\right)(x_{1},x_{2})\nonumber \\
 & =\left(\begin{array}{cc}
0 & 1\\
1 & 0
\end{array}\right)g^{\mu\nu}\delta^{(4)}(x_{1}-x_{2})\nonumber \\
 & +\int dx^{\prime}\left(\begin{array}{ll}
0 & \;\;\Sigma_{A,\rho}^{\mu}\star G_{A}^{\rho\nu}\\
\Sigma_{R,\rho}^{\mu}\star G_{R}^{\rho\nu} & \;\;\Sigma_{C,\rho}^{\mu}\star G_{A}^{\rho\nu}+\Sigma_{R,\rho}^{\mu}\star G_{C}^{\rho\nu}
\end{array}\right)(x_{1},x_{2}),\label{eq:DSE_phy}
\end{alignat}
where we used the shorthand notation $O_{1}\star O_{2}(x_{1},x_{2})\equiv O_{1}(x_{1},x^{\prime})O_{2}(x^{\prime},x_{2})$.
We can assume the system is isotropic, i.e. $G^{\mu\nu}(x_{1},x_{2})=G^{\mu\nu}(x_{1}-x_{2})$,
and the spatial inhomogeneity of the system, as required by the Kubo
formula, is induced by a perturbation. One can obtain the Dyson-Schwinger
equation for retarded and advanced Green's functions in momentum space
(propagators)
\begin{eqnarray}
i\left[g_{\;\rho}^{\mu}(p^{2}-m_{V}^{2})-p^{\mu}p_{\rho}\right]G_{A/R}^{\rho\nu}(p) & = & g^{\mu\nu}+\Sigma_{A/R,\rho}^{\mu}(p)G_{A/R}^{\rho\nu}(p).\label{eq:DSE-RA}
\end{eqnarray}
The free retarded and advanced propagators are given by 
\begin{equation}
G_{(0)A/R}^{\rho\nu}(p)=-i\frac{1}{p^{2}-m^{2}\mp ip_{0}\epsilon}\left(g^{\mu\nu}-\frac{p^{\mu}p^{\nu}}{m_{V}^{2}}\right).
\end{equation}
One can check that $G_{(0)A/R}^{\rho\nu}(p)$ satisfies Eq. (\ref{eq:DSE-RA})
neglecting the last term in the right-hand-side.

% done, QW, 2023.10.2, 16:15

The coupling between the vector meson and quark-antiquark in QGP or
the $\overline{q}qV$ vertex is assumed to be $g_{V}B\overline{\psi}_{q}\gamma^{\mu}\psi_{q}A_{\mu}$
\citep{Manohar:1983md,Fernandez:1993hx,Zacchi:2015lwa,Zacchi:2016tjw}.
Here $B$ denotes the Bethe-Salpeter wave function and can be parameterized
as \citep{Xu:2019ilh,Xu:2021mju}
\begin{eqnarray}
B(p-p^{\prime},p^{\prime}) & = & \frac{1-\exp\left[-(p-2p^{\prime})^{2}/\sigma^{2}\right]}{(p-2p^{\prime})^{2}/\sigma^{2}},
\end{eqnarray}
where $p-p^{\prime}$ and $p^{\prime}$ are momenta of the quark and
anti-quark respectively. We see that the wave function only depends
on the relative momentum.

\begin{figure}
\begin{centering}
\includegraphics[scale=0.2]{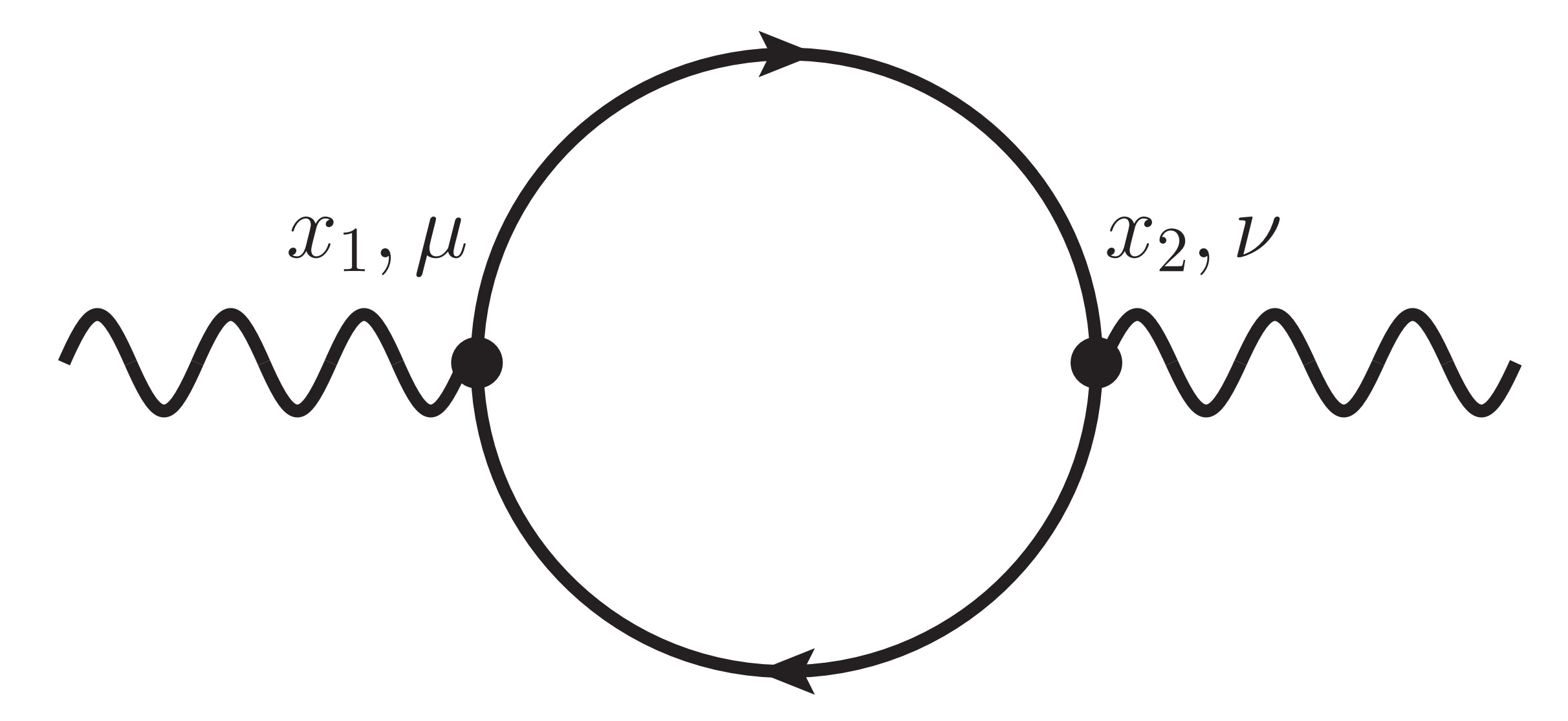}
\par\end{centering}
\caption{The vector meson's self-energy $\Sigma_{\mu\nu}(x_{1},x_{2})$ from
the quark loop (one-loop) contribution. \label{fig:one-loop-self}}
\end{figure}

% done, QW, 2023.10.3, 7:00

We can assume that only when the distance between the quark and anti-quark
is zero can they form a meson, thus we have $1/\sigma\rightarrow0$
and $B=1$. Then the vector meson's self-energy to the lowest order
of the coupling constant $g_{V}$ from the quark one-loop is shown
in Fig. (\ref{fig:one-loop-self}). Applying Eq. (\ref{eq:def_7}),
we can construct retarded and advanced self-energies as 
\begin{eqnarray}
\Sigma_{R}^{\mu\nu}(x_{1},x_{2}) & = & \Sigma_{F}^{\mu\nu}(x_{1},x_{2})-\Sigma_{<}^{\mu\nu}(x_{1},x_{2})\nonumber \\
 & = & g_{V}^{2}\mathrm{Tr}\left[\gamma^{\mu}S_{F}(x_{1},x_{2})\gamma^{\nu}S_{F}(x_{2},x_{1})\right]-g_{V}^{2}\mathrm{Tr}\left[\gamma^{\mu}S_{<}(x_{1},x_{2})\gamma^{\nu}S_{>}(x_{2},x_{1})\right],\nonumber \\
\Sigma_{A}^{\mu\nu}(x_{1},x_{2}) & = & \Sigma_{F}^{\mu\nu}(x_{1},x_{2})-\Sigma_{>}^{\mu\nu}(x_{1},x_{2})\nonumber \\
 & = & g_{V}^{2}\mathrm{Tr}\left[\gamma^{\mu}S_{F}(x_{1},x_{2})\gamma^{\nu}S_{F}(x_{2},x_{1})\right]-g_{V}^{2}\mathrm{Tr}\left[\gamma^{\mu}S_{>}(x_{1},x_{2})\gamma^{\nu}S_{<}(x_{2},x_{1})\right],\label{eq:self-en-r-a-x}
\end{eqnarray}
where $S(x_{1},x_{2})=\left\langle T_{C}\psi(x_{1})\overline{\psi}(x_{2})\right\rangle $
is the two-point Green's function of quarks on the CTP. We have included
a negative sign for the quark loop in Eq. (\ref{eq:self-en-r-a-x}).
Under the assumption that the system is homogeneous in position space,
we obtain self-energies in momentum space 
\begin{eqnarray}
\Sigma_{R}^{\mu\nu}(p) & = & g_{V}^{2}\int\frac{d^{4}k}{(2\pi)^{4}}\left\{ \mathrm{Tr}\left[\gamma^{\mu}S_{F}(k)\gamma^{\nu}S_{F}(k-p)\right]-\mathrm{Tr}\left[\gamma^{\mu}S_{<}(k)\gamma^{\nu}S_{>}(k-p)\right]\right\} .\nonumber \\
\Sigma_{A}^{\mu\nu}(p) & = & g_{V}^{2}\int\frac{d^{4}k}{(2\pi)^{4}}\left\{ \mathrm{Tr}\left[\gamma^{\mu}S_{F}(k)\gamma^{\nu}S_{F}(k-p)\right]+\mathrm{Tr}\left[\gamma^{\mu}S_{>}(k)\gamma^{\nu}S_{<}(k-p)\right]\right\} ,\label{eq:self-energy-1}
\end{eqnarray}
The retarded self-energy is our starting point for derivation of spectral
functions for vector mesons.

% done, QW, 2023.10.3, 7:20

\section{Spectral functions for vector mesons}

\label{sec:Self-Energy-and-Spectral}In this section, we will derive
spectral functions for vector mesons from the retarded self-energy.
We use the CTP formalism in grand-canonical equilibrium which is also
called the real time formalism of the thermal field theory. The vacuum
and thermal equilbrium contributions are incorporated in the same
framework. We assume that quarks and antiquarks are unpolarized and
their distributions are the Fermi-Dirac distribution (\ref{eq:fermi-dirac-dist}). 

% done, QW, 2024.2.14, 8:30

Evaluating the retarded self-energy in Eq. (\ref{eq:self-energy-1})
using the quark propagators in Appendix \ref{sec:quark-propagators},
we obtain 
\begin{eqnarray}
\Sigma_{R}^{\mu\nu}(p) & = & -ig_{V}^{2}\frac{1}{4\pi^{3}}(2I_{1}^{\mu\nu}+I_{2}^{\mu\nu})-ig_{V}^{2}I_{\mathrm{vac}}^{\mu\nu},\label{eq:self-energy}
\end{eqnarray}
where $I_{1}^{\mu\nu}$ and $I_{2}^{\mu\nu}$ are medium parts while
$I_{\mathrm{vac}}^{\mu\nu}$ is the vacuum part. The derivation of
$I_{1}^{\mu\nu}$, $I_{2}^{\mu\nu}$ and $I_{\mathrm{vac}}^{\mu\nu}$
are presented in Appendix \ref{sec:retarded-self-en}. From Eq. (\ref{eq:i1i2-rel})
we have
\begin{align}
\Sigma_{R}^{0i}(p)= & \Sigma_{R}^{i0}(p)=\hat{\mathbf{p}}_{i}\frac{p_{0}}{|\mathbf{p}|}\Sigma_{R}^{00}(p),\nonumber \\
\Sigma_{R}^{ij}(p)= & \hat{\mathbf{p}}_{i}\hat{\mathbf{p}}_{j}\frac{p_{0}^{2}}{|\mathbf{p}|^{2}}\Sigma_{R}^{00}(p)+(\delta^{ij}-\hat{\mathbf{p}}^{i}\hat{\mathbf{p}}^{j})\Sigma_{\perp}(p),
\end{align}
where $\Sigma_{\perp}(p)$ denotes the transverse part of $\Sigma_{R}^{ij}(p)$.
Using above relations, we can greatly simplify the result of full
propagators.

% done, QW, 2023.10.3, 17:00

From Eq. (\ref{eq:quark-propogator}), one can see the only difference
between the retarded and advanced propagators is the sign of the small
positive number $\epsilon$, so the retarded and advanced propagators
or self-energies are complex conjugate to each other, $\Sigma_{A}^{\mu\nu}=-\Sigma_{R}^{\mu\nu*}$
(note that there is an $i$ factor in the definition of the self-energy).
It can be checked that $\Sigma_{R}^{\mu\nu}$ is transverse to $p^{\mu}$
as required by the current conservation. We note that the vacuum contribution
and its real part is divergent and can be renormalized \citep{Gale:1990pn}.
The imaginary part of the vacuum contribution corresponds to the pair
production or annihilation processes.

% done, QW, 2023.10.5, 7:20

Inserting Eq.(\ref{eq:self-energy}) into Eq. (\ref{eq:DSE-RA}) and
introducing 
\begin{equation}
\widetilde{\Sigma}_{R}^{\mu\nu}(p)=-i\Sigma_{R}^{\mu\nu}(p)=-g_{V}^{2}\frac{1}{4\pi^{3}}(2I_{1}^{\mu\nu}+I_{2}^{\mu\nu})-g_{V}^{2}I_{\mathrm{vac}}^{\mu\nu},
\end{equation}
we obtain 
\begin{equation}
\left[G_{R}^{-1}(p)\right]^{\mu\nu}=i\left[g^{\mu\nu}(p^{2}-m_{V}^{2})-p^{\mu}p^{\nu}-\widetilde{\Sigma}_{R}^{\mu\nu}(p)\right].\label{eq:DES-inverse}
\end{equation}
From the definition of $I_{1,2}^{\mu\nu}$, we find that they are
written in terms of projectors related to three momentum $\mathbf{p}$.
Therefore, we assume $G_{R}^{\mu\nu}$ has the same structure with
$\Sigma_{R}^{\mu\nu}$ and can be written as
\begin{eqnarray}
G_{R}^{00} & = & iA,\nonumber \\
G_{R}^{0i} & = & G_{R}^{i0}=i\hat{\mathbf{p}}_{i}B,\nonumber \\
G_{R}^{ij} & = & i\left[(\delta_{ij}-\hat{\mathbf{p}}_{i}\hat{\mathbf{p}}_{j})C+\hat{\mathbf{p}}_{i}\hat{\mathbf{p}}_{j}D\right],\label{eq:G_R}
\end{eqnarray}
where $A,B,C,D$ are functions of $p$ and are not independent since
$G_{\lessgtr}^{\mu\nu}$ are transverse to $p^{\mu}$. By solving
$(G_{R}^{-1})_{\;\rho}^{\mu}G_{R}^{\rho\nu}=g^{\mu\nu}$, we find
\begin{eqnarray}
A & = & \frac{1}{m^{2}}\frac{p_{0}^{2}-m^{2}+(p_{0}^{2}/|\mathbf{p}|^{2})\widetilde{\Sigma}_{R}^{00}}{p^{2}-m^{2}+(p^{2}/|\mathbf{p}|^{2})\widetilde{\Sigma}_{R}^{00}},\nonumber \\
C & = & \frac{1}{p^{2}-m^{2}+\widetilde{\Sigma}_{\perp}},
\end{eqnarray}
where $\widetilde{\Sigma}_{\perp}=-i\Sigma_{\perp}$. Other two functions
$B$ and $D$ can be expressed in terms of $A$ and will be discussed
later. We can also define $\widetilde{G}_{R,A}^{\mu\nu}(p)=-iG_{R,A}^{\mu\nu}(p)$
to remove the factor $i$ in the definition of $G_{R,A}^{\mu\nu}(p)$.
The advanced full propagator $\widetilde{G}_{A}^{\mu\nu}$ can be
obtained by $\widetilde{G}_{A}^{\mu\nu}=\widetilde{G}_{R}^{\mu\nu*}$.
It should be emphasized that this relation holds only for an unpolarized
case.

We can construct $G_{<}^{\mu\nu}$ from $G_{A}^{\mu\nu}$ and $G_{R}^{\mu\nu}$
as \citep{fetter2003quantum,zubarev1997statistical}
\begin{eqnarray}
G_{<}^{\mu\nu}(p) & = & in_{B}(p_{0})\left[\widetilde{G}_{R}^{\mu\nu}(p)-\widetilde{G}_{A}^{\mu\nu}(p)\right]\nonumber \\
 & = & -2n_{B}(p_{0})\mathrm{Im}\widetilde{G}_{R}^{\mu\nu}(p),\label{eq:less_R}
\end{eqnarray}
where $n_{B}(p_{0})=1/(e^{\beta p_{0}-\beta\mu_{V}}-1)$ is the Bose-Einstein
distribution with the inverse temperature $\beta\equiv1/T$ and the
vecctor meson's chemical potential $\mu_{V}$ ($\mu_{V}=0$ for the
unflavored meson). Note that there is an $i$ factor in the definition
of the propagator without tilde. From Eq. (\ref{eq:less_R}), we find
the real part of $A,B,C,D$ have no contributions to the spectral
function, and the imaginary part of $A,B,D$ have following constraints
from $p_{\mu}G_{<}^{\mu\nu}=0$, 
\begin{eqnarray}
p_{0}\mathrm{Im}A-|\mathbf{p}|\mathrm{Im}B & = & 0,\nonumber \\
p_{0}\mathrm{Im}B-|\mathbf{p}|\mathrm{Im}D & = & 0.
\end{eqnarray}
Inserting Eq. (\ref{eq:G_R}) into Eq. (\ref{eq:less_R}), one can
obtain 
\begin{eqnarray}
G_{<}^{\mu\nu}(p) & = & -2n_{B}(p_{0})\left[\Delta_{T}^{\mu\nu}\rho_{T}(p)+\Delta_{L}^{\mu\nu}\rho_{L}(p)\right],\label{eq:G_less(0)}
\end{eqnarray}
or equivalently 
\begin{equation}
\mathrm{Im}\widetilde{G}_{R}^{\mu\nu}(p)=\Delta_{T}^{\mu\nu}\rho_{T}(p)+\Delta_{L}^{\mu\nu}\rho_{L}(p).\label{eq:im-gr}
\end{equation}
In Eqs. (\ref{eq:G_less(0)},\ref{eq:im-gr}), we defined 
\begin{align}
\Delta_{T}^{\mu\nu}= & -g^{\mu0}g^{\nu0}+g^{\mu\nu}+\frac{\mathbf{p}^{\mu}\mathbf{p}^{\nu}}{|\mathbf{p}|^{2}},\nonumber \\
\Delta_{L}^{\mu\nu}= & \Delta^{\mu\nu}-\Delta_{T}^{\mu\nu}\equiv g^{\mu\nu}-\frac{p^{\mu}p^{\nu}}{p^{2}}-\Delta_{T}^{\mu\nu},\label{eq:projectors}
\end{align}
as the transverse and longitudinal projector respectively with $\mathbf{p}^{\mu}=(0,\mathbf{p})$,
and $\rho_{T,L}$ are spectral functions in the transverse and longitudinal
directions given by 
\begin{eqnarray}
\rho_{T}(p) & = & -\mathrm{Im}C=-\mathrm{Im}\frac{1}{p^{2}-m^{2}+\widetilde{\Sigma}_{\perp}(p)+i\,\mathrm{sgn}(p_{0})\varepsilon},\nonumber \\
\rho_{L}(p) & = & -\frac{p^{2}}{|\mathbf{p}|^{2}}\mathrm{Im}A=-\mathrm{Im}\frac{1}{p^{2}-m^{2}+\frac{p^{2}}{|\mathbf{p}|^{2}}\widetilde{\Sigma}_{00}(p)+i\,\mathrm{sgn}(p_{0})\varepsilon},\label{eq:rho-tl}
\end{eqnarray}
where $\widetilde{\Sigma}_{\perp}$ and $\widetilde{\Sigma}_{00}$
are from $\widetilde{\Sigma}_{R}^{\mu\nu}$: $\widetilde{\Sigma}_{\perp}\equiv-(1/2)\Delta_{\mu\nu}^{T}\widetilde{\Sigma}_{R}^{\mu\nu}$
and $\widetilde{\Sigma}_{00}=\widetilde{\Sigma}_{R}^{00}$, $\mathrm{sgn}(p_{0})$
is the sign of $p_{0}$, and $\varepsilon$ is an infinitesimal positive
number. One can check in Eq. (\ref{eq:projectors}) that $p_{\mu}\Delta_{T}^{\mu\nu}=p_{\mu}\Delta_{L}^{\mu\nu}=0$.
In Eq. (\ref{eq:rho-tl}), one can verify that the real parts of $\widetilde{\Sigma}_{\perp}$
and $\widetilde{\Sigma}_{00}$ contribute to the mass correction while
the imaginary parts of $\widetilde{\Sigma}_{\perp}$ and $\widetilde{\Sigma}_{00}$
determines the width or life-time of the quasi-particle mode. For
free vector mesons, the spectral functions are $\rho_{T}^{(0)}=\rho_{L}^{(0)}=\pi\mathrm{sgn}(p_{0})\delta(p^{2}-m^{2})$,
which give $G_{<}^{\mu\nu}(p)$ for the free vector meson following
Eq. (\ref{eq:G_less(0)}) and $\mathrm{Im}\widetilde{G}_{R}^{\mu\nu}(p)$
for the free vector meson following Eq. (\ref{eq:im-gr}).

\section{Kubo formula in linear response theory}

\label{sec:Kubo-Formula}In this section we use the Kubo formula in
linear response theory to calculate the non-equilibrium correction
to $G_{<}^{\mu\nu}(p)$. The Kubo formula has been derived in Zubarev's
approach to non-equilibrium density operator \citep{Zubarev_1979,Hosoya:1983id,Becattini:2019dxo}.

% done, QW, 2024.2.14, 10:30

According to the Kubo formula, the linear response of the expectation
value of an operator $\hat{O}$ to the perturbation $\partial_{\mu}\beta_{\nu}$
has the form
\begin{align}
\left\langle \hat{O}(x)\right\rangle = & \left\langle \hat{O}\right\rangle _{\mathrm{LE}}+\partial_{\mu}\beta_{\nu}(x)\lim_{K^{\mu}\rightarrow0}\frac{\partial}{\partial K_{0}}\nonumber \\
 & \times\mathrm{Im}\left[iT(x)\int_{-\infty}^{t}d^{4}x^{\prime}\left\langle \left[\hat{O}(x),\hat{T}^{\mu\nu}(x^{\prime})\right]\right\rangle _{\mathrm{LE}}e^{-iK\cdot(x^{\prime}-x)}\right],\label{eq:linear-response}
\end{align}
where $\left\langle \hat{O}(x)\right\rangle \equiv\mathrm{Tr}\left[\hat{\rho}\hat{O}(x)\right]$
and $\left\langle \hat{O}(x)\right\rangle _{\mathrm{LE}}\equiv\mathrm{Tr}\left[\hat{\rho}_{\mathrm{LE}}\hat{O}(x)\right]$
with $\hat{\rho}$ and $\hat{\rho}_{\mathrm{LE}}$ being the non-equilibrium
and local equilibrium density operator respectively \citep{Becattini:2019dxo},
$\beta^{\mu}(x)\equiv u^{\mu}(x)/T(x)$ with $u^{\mu}(x)$ and $T(x)$
being the local velocity and temperature respectively, $K^{\mu}$
is the momentum roughly equals to $\pi/L$ with $L$ being the length
of the system, and 
\begin{equation}
\hat{T}^{\mu\nu}=\hat{F}_{\;\;\alpha}^{\mu}\hat{F}^{\alpha\nu}+m_{V}^{2}\hat{A}^{\mu}\hat{A}^{\nu}-g^{\mu\nu}\left(-\frac{1}{4}\hat{F}_{\rho\eta}\hat{F}^{\rho\eta}+\frac{1}{2}m_{V}^{2}\hat{A}_{\rho}\hat{A}^{\rho}\right),\label{eq:stress-tensor-op}
\end{equation}
is the energy-momentum tensor for the vector field. Detailed derivation
of Eq. (\ref{eq:linear-response}) is given in Ref. \citep{Becattini:2019dxo}.

% done, QW, 2024.2.14, 10:30

Now we set $\hat{O}(x)$ to be the operator corresponding to $G_{<}^{\mu\nu}$
\begin{eqnarray}
\hat{G}_{<}^{\mu\nu}(x,p) & = & \int d^{4}ye^{ip\cdot y}\hat{A}^{\nu}\left(x-\frac{y}{2}\right)\hat{A}^{\mu}\left(x+\frac{y}{2}\right),\label{eq:G-less-operator}
\end{eqnarray}
which gives $G_{<}^{\mu\nu}=\left\langle \hat{G}_{<}^{\mu\nu}(x,p)\right\rangle $.
In Eqs. (\ref{eq:stress-tensor-op}) and (\ref{eq:G-less-operator})
we explicitly show the ``hat'' on the field operator $\hat{A}^{\mu}$
which we have suppressed in Sec. \ref{sec:green-functions} and \ref{sec:wigner-functions}
just to emphasize their operator's nature in the Kubo formula (\ref{eq:linear-response}).
When inserting Eqs. (\ref{eq:stress-tensor-op}) and (\ref{eq:G-less-operator})
into Eq. (\ref{eq:linear-response}), the vector field $\hat{A}^{\mu}$
can be approximated as the free field at the leading order in space-time
gradient, since $\partial_{\mu}\beta_{\nu}(x)$ is already of the
next-to-leading order. 

% done, QW, 2024.2.14, 10:30

Substituting $\hat{G}_{<}^{\mu\nu}(x,p)$ in (\ref{eq:G-less-operator})
into Eq.(\ref{eq:linear-response}), one obtains the next-to-leading
order term of $G_{<}^{\mu\nu}$ as
\begin{eqnarray}
\delta G_{<}^{\mu\nu}(x,p) & \equiv & \left\langle \hat{G}_{<}^{\mu\nu}(x,p)\right\rangle -\left\langle \hat{G}_{<}^{\mu\nu}(x,p)\right\rangle _{\mathrm{LE}}\nonumber \\
 & = & 4T\lim_{K^{\mu}\rightarrow0}\frac{\partial}{\partial K_{0}}\mathrm{Im}\int\frac{dp_{1}^{0}dp_{2}^{0}}{2\pi}\frac{n_{B}(p_{1}^{0})-n_{B}(p_{2}^{0})}{p_{1}^{0}-p_{2}^{0}+K^{0}+i\epsilon}\nonumber \\
 &  & \times\delta(p^{0}-\frac{p_{1}^{0}+p_{2}^{0}}{2})\partial_{\gamma}\beta_{\lambda}(x)\sum_{a,b=L,T}\rho_{a}(p_{1})\rho_{b}(p_{2})I_{ab}^{\mu\nu\gamma\lambda}(p_{1},p_{2})\label{eq:G_<(1)}
\end{eqnarray}
where $p_{1}=(p_{1}^{0},\mathbf{p}-\mathbf{K}/2),p_{2}=(p_{2}^{0},\mathbf{p}+\mathbf{K}/2)$,
$n_{B}(p_{0})$ is the Bose-Einstein distribution defined after Eq.
(\ref{eq:less_R}), and $\rho_{L,T}$ are given in Eq. (\ref{eq:rho-tl}).
Note that integral ranges for $p_{1,2}^{0}$ are different from Ref.
\citep{Li:2022vmb}. The tensor $I_{ab}^{\mu\nu\gamma\lambda}(p_{1},p_{2})$
can be expressed in terms of projectors $\Delta_{L,T}^{\mu\nu}$ as
\begin{eqnarray}
I_{ab}^{\mu\nu\gamma\lambda}(p_{1},p_{2}) & = & (p_{1}^{\lambda}p_{2}^{\gamma}+p_{1}^{\gamma}p_{2}^{\lambda})\Delta_{a,\alpha}^{\nu}(p_{1})\Delta_{b}^{\mu\alpha}(p_{2})\nonumber \\
 &  & +(p_{1,\alpha}p_{2}^{\alpha}-m_{V}^{2})\left[\Delta_{a}^{\gamma\nu}(p_{1})\Delta_{b}^{\mu\lambda}(p_{2})+\Delta_{a}^{\lambda\nu}(p_{1})\Delta_{b}^{\mu\gamma}(p_{2})\right]\nonumber \\
 &  & -\left[p_{1}^{\gamma}p_{2}^{\alpha}\Delta_{a,\alpha}^{\nu}(p_{1})\Delta_{b}^{\mu\lambda}(p_{2})+p_{2}^{\gamma}p_{1}^{\alpha}\Delta_{a}^{\lambda\nu}(p_{1})\Delta_{b,\alpha}^{\mu}(p_{2})\right]\nonumber \\
 &  & -\left[p_{1,\alpha}p_{2}^{\lambda}\Delta_{a}^{\gamma\nu}(p_{1})\Delta_{b}^{\mu\alpha}(p_{2})+p_{1}^{\lambda}p_{2,\alpha}\Delta_{a}^{\alpha\nu}(p_{1})\Delta_{b}^{\mu\gamma}(p_{2})\right]\nonumber \\
 &  & -g^{\gamma\lambda}\left[g_{\beta\alpha}(p_{2,\rho}p_{1}^{\rho}-m_{V}^{2})-p_{1,\beta}p_{2,\alpha}\right]\Delta_{a}^{\alpha\nu}(p_{1})\Delta_{b}^{\mu\beta}(p_{2}).\label{eq:I_ab}
\end{eqnarray}
Then we integrate Eq. (\ref{eq:G_<(1)}) over $p_{0}$ from $0$ to
$+\infty$ to exclude the contribution from anti-particles. As we
have mentioned above, the limit $K^{\mu}\rightarrow0$ should be taken
in the last step, thus the integral of Eq. (\ref{eq:G_<(1)}) can
be simplified as 
\begin{equation}
\int_{0}^{+\infty}dp_{0}\delta G_{<}^{\mu\nu}(x,p)\approx2T\xi_{\gamma\lambda}\int_{0}^{\infty}dp_{1}^{0}\frac{\partial n_{B}(p_{1}^{0})}{\partial p_{1}^{0}}\sum_{a,b=L,T}\rho_{a}(p_{1}^{0},\mathbf{p})\rho_{b}(p_{1}^{0},\mathbf{p})I_{ab}^{\mu\nu\gamma\lambda}(p_{1}^{0},\mathbf{p},p_{1}^{0},\mathbf{p}),\label{eq:G_<(1)_p0_integral}
\end{equation}
where $\xi_{\gamma\lambda}=\partial_{(\gamma}\beta_{\lambda)}$ denotes
the thermal shear tensor.

The spin alignment coupled with the thermal shear tensor is given
by 
\begin{eqnarray}
\delta\rho_{00} & = & \frac{L^{\mu\nu}(p_{\mathrm{on}})\int_{0}^{+\infty}dp_{0}\left[G_{\mu\nu}^{<}(x,p)+\delta G_{\mu\nu}^{<}(x,p)\right]}{-\Delta^{\mu\nu}(p_{\mathrm{on}})\int_{0}^{+\infty}dp_{0}\left[G_{\mu\nu}^{<}(x,p)+\delta G_{\mu\nu}^{<}(x,p)\right]},\label{eq:dev-rho00}
\end{eqnarray}
where $G_{\mu\nu}^{<}(x,p)$ is given in Eq. (\ref{eq:G_less(0)})
while $\delta G_{\mu\nu}^{<}(x,p)$ is given in (\ref{eq:G_<(1)_p0_integral}),
and $L^{\mu\nu}(p_{\mathrm{on}})$ is defined in Eq. (\ref{eq:l-munu-on}).
The above formula is the starting point for us to evaluate the correction
to $\rho_{00}$ from the shear stress tensor in the next section.

We should note about the difference between the average taken in $G_{\mu\nu}^{<}(x,p)$
given by Eq. (\ref{eq:G_less(0)}) {[}as well as other avarages in
Sec. \ref{sec:green-functions}{]} and the one taken in $\left\langle \hat{G}_{<}^{\mu\nu}(x,p)\right\rangle $
in Eq. (\ref{eq:G_<(1)}). The local equilibrium average is implied
for the former, while the non-equilibrium average is implied for the
latter. For notational simplicity, we do not put ``LE'' index to
local equilibrium averages in this paper except in the Kubo formula
Eqs. (\ref{eq:linear-response}) and (\ref{eq:G_<(1)}).

% done, QW, 2023.2.14, 10:40

\section{Spin alignment correction from shear tensor}

\label{sec:Self-Energy-and-Shear}In this section, we will calculate
the spin alignment correction from the shear tensor. To this end,
we adopt two approximations to evaluate self-energies and spectral
functions of unflavored vector mesons: the HTL and quasi-particle
approximation.

\subsection{HTL approximation}

Under the HTL approximation, the external momentum of the vector meson's
self-energy is of order $g_{V}T$ which is called ``soft'' while
the quark loop momentum is of order $T$ which is called ``hard''
\citep{Weldon:1982aq,Pisarski:1988vd,Braaten:1989mz,Thoma:2000dc,Blaizot:2001nr,Bellac:2011kqa}.
This condition is not satisfied for the real vector meson in the thermal
environment at RHIC and LHC with $p_{0}>m_{V}\gg T$. The reason that
we still consider the HTL approximation is that the self-energies
in this approximation is analytical and the calculation of the spin
density matrix is transparent. In other words, we treat the HTL approximation
as a toy model to show the underlying physics.

We can consider massless quarks for simplicity. Note that the vacuum
term is not included since the imaginary part of the vacuum term corresponds
to the process that one particle decomposes into two on-shell quarks,
i.e. $p^{0}>k^{0}$, which is beyond the HTL approximation. The vacuum
contribution is proportional to $p^{2}\Delta^{\mu\nu}$ as required
by the Ward identity, which is of order $g_{V}^{4}T^{2}$ since $p\sim g_{V}T$.
The self-energy in the HTL approximation reads 
\begin{eqnarray}
\widetilde{\Sigma}_{00}(p) & = & 3m_{T}^{2}\left(1-\frac{p_{0}}{2|\mathbf{p}|}\ln\frac{p_{0}+|\mathbf{p}|+i\epsilon}{p_{0}-|\mathbf{p}|+i\epsilon}\right),\nonumber \\
\widetilde{\Sigma}_{\perp}(p) & = & -\frac{3}{2}m_{T}^{2}\frac{p_{0}^{2}}{|\mathbf{p}|^{2}}\left(1-\frac{p_{0}^{2}-|\mathbf{p}|^{2}}{2p_{0}|\mathbf{p}|}\ln\frac{p_{0}+|\mathbf{p}|+i\epsilon}{p_{0}-|\mathbf{p}|+i\epsilon}\right),\label{eq:HTL-self-energy-00-perp}
\end{eqnarray}
where $m_{T}^{2}=g_{V}^{2}T^{2}/9$ denotes the thermal mass. The
real and imaginary parts of $\widetilde{\Sigma}^{00}$ and $\widetilde{\Sigma}_{\perp}$
can be obtained as 
\begin{eqnarray}
\mathrm{Re}\widetilde{\Sigma}_{00}(p) & = & 3m_{T}^{2}\left(1-\frac{p_{0}}{2|\mathbf{p}|}\ln\left|\frac{p_{0}+|\mathbf{p}|}{p_{0}-|\mathbf{p}|}\right|\right),\nonumber \\
\mathrm{Re}\widetilde{\Sigma}_{\perp}(p) & = & -\frac{3}{2}m_{T}^{2}\frac{p_{0}^{2}}{|\mathbf{p}|^{2}}\left(1-\frac{p_{0}^{2}-|\mathbf{p}|^{2}}{2p_{0}|\mathbf{p}|}\ln\left|\frac{p_{0}+|\mathbf{p}|}{p_{0}-|\mathbf{p}|}\right|\right),\nonumber \\
\mathrm{Im}\widetilde{\Sigma}_{00}(p) & = & \pi\frac{3}{2}m_{T}^{2}\frac{p_{0}}{|\mathbf{p}|}\theta(|\mathbf{p}|^{2}-p_{0}^{2}),\nonumber \\
\mathrm{Im}\widetilde{\Sigma}_{\perp}(p) & = & -\pi\frac{3}{4}m_{T}^{2}\frac{p_{0}(p_{0}^{2}-|\mathbf{p}|^{2})}{|\mathbf{p}|^{3}}\theta(|\mathbf{p}|^{2}-p_{0}^{2}),\label{eq:real-Im-HTL-self-energy}
\end{eqnarray}
where $\theta(x)$ is the Heaviside step function. We see that the
imaginary parts are non-vanishing only in space-like region of $p$.

% done, QW, 2023.10.15, 9:00

Under the HTL approximation, one can get the inequality $p_{0}\sim m_{T}\sim g_{V}T\ll m_{V}$,
which provides a natural power counting in $\alpha\equiv m_{T}/m_{V}$.
We also assume $p_{0}^{2}<|\mathbf{p}|^{2}$, so there is no pole
contribution. Then the spectral function $\rho_{L/T}$ in (\ref{eq:rho-tl})
can be approximated as 
\begin{eqnarray}
\rho_{T}(p) & = & \frac{\mathrm{Im}\widetilde{\Sigma}_{\perp}(p)}{m_{V}^{4}}+\mathcal{O}(\alpha^{5})\nonumber \\
 & = & -\pi\frac{3}{4}\frac{m_{T}^{2}}{m_{V}^{4}}\frac{p_{0}p^{2}}{|\mathbf{p}|^{3}}\theta(|\mathbf{p}|^{2}-p_{0}^{2})+\frac{1}{m_{V}^{2}}\mathcal{O}(\alpha^{3}),\nonumber \\
\rho_{L}(p) & = & \frac{p^{2}}{|\mathbf{p}|^{2}}\frac{\mathrm{Im}\widetilde{\Sigma}^{00}(p)}{m_{V}^{4}}+\mathcal{O}(\alpha^{5})\nonumber \\
 & = & \pi\frac{3}{2}\frac{m_{T}^{2}}{m_{V}^{4}}\frac{p_{0}p^{2}}{|\mathbf{p}|^{3}}\theta(|\mathbf{p}|^{2}-p_{0}^{2})+\frac{1}{m_{V}^{2}}\mathcal{O}(\alpha^{3}).\label{eq:rho-tl-expansion}
\end{eqnarray}
for $p^{0}\ll m_{V}$. Using Eq. (\ref{eq:rho-tl-expansion}) in Eq.
(\ref{eq:dev-rho00}), we can get the leading order term of $\delta\rho_{00}$
\begin{eqnarray}
\delta\rho_{00}^{(0)} & \approx & -T\frac{\xi_{\gamma\lambda}\int_{0}^{|\mathbf{p}|}dp_{0}\frac{\partial n(p_{0})}{\partial p_{0}}\rho_{a}(p)\rho_{b}(p)\left[I_{ab}^{22\gamma\lambda}(p)-\frac{1}{3}I_{ab}^{ii\gamma\lambda}(p)\right]}{(g_{\mu0}g_{\nu0}-g_{\mu\nu})\int_{0}^{|\mathbf{p}|}dp_{0}n(p_{0})\left[\Delta_{L}^{\mu\nu}\rho_{L}(p)+\Delta_{T}^{\mu\nu}\rho_{T}(p)\right]}\nonumber \\
 &  & +\frac{\int_{0}^{\infty}dp^{0}n_{B}(p_{0})\left[(\Delta_{T}^{22}-\frac{1}{3}\Delta_{T}^{ii})\rho_{T}(p)+(\Delta_{L}^{22}-\frac{1}{3}\Delta_{L}^{ii})\rho_{L}(p)\right]}{\left\{ \int_{0}^{|\mathbf{p}|}dp^{0}n_{B}(p_{0})\left[\Delta_{T}^{kk}\rho_{T}(p)+\Delta_{L}^{kk}\rho_{L}(p)\right]\right\} ^{2}}\nonumber \\
 &  & \times T\xi_{\gamma\lambda}\int_{0}^{|\mathbf{p}|}dp^{0}\frac{\partial n_{B}(p_{0})}{\partial p_{0}}\sum_{a,b=T,L}\rho_{a}(p)\rho_{b}(p)I_{ab}^{jj\gamma\lambda}(p;p),\label{eq:leading-rho00}
\end{eqnarray}
where $\Delta_{L,T}^{\mu\nu}$ are projectors defined in Eq. (\ref{eq:projectors}).
In Eq. (\ref{eq:leading-rho00}), the polarization vector can be approximated
as $\epsilon^{\mu}(0,p_{\mathrm{on}})=(0,0,1,0)+\mathcal{O}(\alpha)$
with $|\mathbf{p}|\ll E_{p}\approx m_{V}$, where we choose $y$ direction
as the spin quantization direction. So $\Delta^{\mu\nu}(p_{\mathrm{on}})$
can be approximated as $g^{\mu\nu}-g^{\mu0}g^{\nu0}$. The leading
order $I_{ab,(0)}^{\mu\nu\gamma\lambda}$ is given by 
\begin{eqnarray}
I_{ab}^{\mu\nu\gamma\lambda}(p) & \approx & -m_{V}^{2}\left[\Delta_{a}^{\gamma\nu}(p)\Delta_{b}^{\mu\lambda}(p)+\Delta_{a}^{\lambda\nu}(p)\Delta_{b}^{\mu\gamma}(p)\right]\nonumber \\
 &  & +m_{V}^{2}\eta^{\gamma\lambda}\eta_{\beta\alpha}\Delta_{a}^{\alpha\nu}(p)\Delta_{b}^{\mu\beta}(p),\label{eq:I_ab_(0)}
\end{eqnarray}
which is $\mathcal{O}(m_{V}^{2})$. Finally we can estimate 
\begin{equation}
\delta\rho_{00}^{(0)}\sim\frac{m_{V}^{-2}\alpha^{2}\times m_{V}^{-2}\alpha^{2}\times m_{V}^{2}}{m_{V}^{-2}\alpha^{2}}\times\xi\sim\alpha^{2}\xi,
\end{equation}
where $\xi\equiv|\xi_{\gamma\lambda}|$ is the magnitude of the thermal
shear tensor. If we set the parameters' values as $g_{V}=1$, $T=$150
MeV, $m_{V}=$1020 MeV, the coupling between the spin alignment and
the shear tensor is about $\alpha^{2}\sim\mathcal{O}(10^{-2})$. If
we further use\textcolor{red}{{} }$\xi\sim0.01$, then we obtain $\delta\rho_{00}^{(0)}\sim\mathcal{O}(10^{-4})$,
which is much smaller than the contribution from the coalescence model
via strong force fields \citep{Sheng:2022wsy}.

% done, QW, 2023.11.7, 16:00

\subsection{Vector mesons as resonances}

Now we consider the realistic case that vector mesons are resonances
so that the coalescence and dissociation processes can happen. In
this case, we have $p^{0}>E_{k}$ and $(p-k)^{2}>0$. The small imaginary
numbers in the quark loop integral become $\pm i(E_{k}\pm p_{0})\epsilon\propto i\epsilon$
in $J_{\pm}(p;n_{1},n_{2})$ in Eq. (\ref{eq:jn1n2}). Therefore,
the vector meson's self-energies read 
\begin{eqnarray}
\widetilde{\Sigma}_{00}(p) & = & -g_{V}^{2}\frac{1}{4\pi^{2}}\left\{ -4J_{0}(-1,2)+2\frac{p_{0}}{|\mathbf{p}|}\left[J_{+}(p;0,1)-J_{-}(p;0,1)\right]\right.\nonumber \\
 &  & +\frac{p^{2}}{2|\mathbf{p}|}\left[J_{+}(p;-1,1)+J_{-}(p;-1,1)\right]\nonumber \\
 &  & \left.+\frac{2}{|\mathbf{p}|}\left[J_{+}(p,1,1)+J_{-}(p,1,1)\right]\right\} -g_{V}^{2}I_{\mathrm{vac}}^{00},\nonumber \\
\widetilde{\Sigma}_{\perp}(p) & = & -g_{V}^{2}\frac{1}{16\pi^{2}}\frac{1}{|\mathbf{p}|^{3}}\left\{ (8p^{2}|\mathbf{p}|+16|\mathbf{p}|^{3})J_{0}(-1,2)\right.\nonumber \\
 &  & -(p^{4}+2p^{2}|\mathbf{p}|^{2})\left[J_{+}(p;-1,1)+J_{-}(p;-1,1)\right]\nonumber \\
 &  & -4p_{0}^{2}\left[J_{+}(p;1,1)+J_{-}(p;1,1)\right]-4p_{0}p^{2}\left[J_{+}(p;0,1)-J_{-}(p;0,1)\right]\nonumber \\
 &  & \left.+4|\mathbf{p}|^{2}\left[J_{+}(p;-1,3)+J_{-}(p;-1,3)\right]\right\} -g_{V}^{2}I_{\mathrm{vac}}^{\perp},
\end{eqnarray}
where $I_{\mathrm{vac}}^{\perp}\equiv(1/2)(\delta_{ij}-\hat{\mathbf{p}}_{i}\hat{\mathbf{p}}_{j})I_{\mathrm{vac}}^{ij}$.
Note that the vacuum contributions to real parts of self-energies
are canceled by renormalization. When evaluating imaginary parts,
we note that $J_{0}(n_{1},n_{2})$ is real and $\mathrm{Im}J_{+}$
is non-zero in the region $p^{2}+2p_{0}E_{k}<2|\mathbf{k}||\mathbf{p}|$,
which cannot be satisfied under the quasi-particle approximation with
$p_{0}>|\mathbf{p}|$ and $E_{k}>|\mathbf{k}|$. So the imaginary
parts come from $\mathrm{Im}J_{-}(p;n_{1},n_{2})$ within the range
$-2|\mathbf{k}||\mathbf{p}|\le p^{2}-2p_{0}E_{k}\le2|\mathbf{k}||\mathbf{p}|$
as 
\begin{align}
\mathrm{Im}J_{-}(p;n_{1},n_{2})= & -\pi\int_{E_{\mathrm{min}}}^{E_{\mathrm{max}}}dE_{k}(E_{k}^{2}-m_{q}^{2})^{(n_{2}-1)/2}E_{k}^{n_{1}+1}\nonumber \\
 & \times\left[\frac{1}{e^{(E_{k}-\mu_{q})/T}+1}+\frac{1}{e^{(E_{k}+\mu_{q})/T}+1}\right],
\end{align}
where $\mu_{q}$ is the chemical potential of quarks, $m_{q}$ is
the quark mass, and $E_{\mathrm{max/min}}$ is 
\begin{equation}
E_{\mathrm{max/min}}=\sqrt{\left(\pm\frac{\mathbf{p}}{2}+\frac{p_{0}}{2}\sqrt{1-\frac{4m_{q}^{2}}{p^{2}}}\right)^{2}+m_{q}^{2}}.
\end{equation}
We see that imaginary parts exist only when $p^{2}>4m_{q}^{2}$. Then
the imaginary parts of self-energies read 
\begin{eqnarray}
\mathrm{Im}\widetilde{\Sigma}_{00}(p) & = & -g_{V}^{2}\frac{1}{2\pi^{2}|\mathbf{p}|}\mathrm{Im}\left[J_{-}(p;1,1)-p_{0}J_{-}(p;0,1)+\frac{p^{2}}{4}J_{-}(p;-1,1)\right]\nonumber \\
 &  & -g_{V}^{2}\mathrm{Im}I_{\mathrm{vac}}^{00},\nonumber \\
\mathrm{Im}\widetilde{\Sigma}_{\perp}(p) & = & -g_{V}^{2}\frac{1}{16\pi^{2}|\mathbf{p}|^{3}}\mathrm{Im}\left[-(p^{4}+2|\mathbf{p}|^{2}p^{2})J_{-}(p;-1,1)-4p_{0}^{2}J_{-}(p;1,1)\right.\nonumber \\
 &  & \left.+4p_{0}p^{2}J_{-}(p;0,1)+4|\mathbf{p}|^{2}J_{-}(p;-1,3)\right]-g_{V}^{2}\mathrm{Im}I_{\mathrm{vac}}^{\perp}.\label{eq:Im-self-energy-00-perp}
\end{eqnarray}
Note that vacuum contributions are included in imaginary parts of
self-energies, which correspond to pair production and annihilation
(dissociation and combination) processes involving on-shell particles
in the initial and final states (the meson, quark and antiquark are
all on-shell).

% done, QW, 2023.11.8, 14:00

\subsubsection{Quasi-particle approximation \label{subsec:qpa}}

We take the quasi-particle approximation (QPA) for the vector meson
that $g_{V}$ is not very large and the self-energies are assumed
to be small compared with $m_{V}^{2}$. In this case, the spectral
functions in Eq. (\ref{eq:rho-tl}) have narrow peaks around $E_{p}^{V}$.
In the region near $p_{0}=E_{p}^{V}$, we can approximate the self-energies
as their on-shell values, i.e., $\widetilde{\Sigma}_{00}(p)\approx\widetilde{\Sigma}_{00}(p_{\mathrm{on}})$
and $\widetilde{\Sigma}_{\perp}(p)\approx\widetilde{\Sigma}_{\perp}(p_{\mathrm{on}})$.
Then spectral functions for transverse/longitudinal modes can be approximated
as 
\begin{align}
\rho_{T/L}(p)= & \rho_{T/L}^{\mathrm{pole}}(p)+\rho_{T/L}^{\mathrm{cut}}(p)\nonumber \\
\approx & \pi\,\mathrm{sgn}(p_{0})\delta\left[p_{0}^{2}-(E_{p}^{V}+\Delta E_{T/L})^{2}\right]\theta\left(4m_{q}^{2}-p^{2}\right)\nonumber \\
 & +\frac{m_{V}\Gamma_{T/L}}{\left[p_{0}^{2}-(E_{p}^{V}+\Delta E_{T/L})^{2}\right]^{2}+m_{V}^{2}\Gamma_{T/L}^{2}}\theta\left(p^{2}-4m_{q}^{2}\right),\label{eq:Breit-Wigner}
\end{align}
where $\Gamma_{T/L}$ are widths and $\Delta E_{T/L}$ are energy
shifts for transverse/longitudinal modes approximated as 
\begin{align}
\Gamma_{T}= & \frac{1}{m_{V}}\mathrm{Im}\widetilde{\Sigma}_{\perp}(p_{\mathrm{on}}),\nonumber \\
\Gamma_{L}= & \frac{m_{V}}{|\mathbf{p}|^{2}}\mathrm{Im}\widetilde{\Sigma}^{00}(p_{\mathrm{on}}),\nonumber \\
\Delta E_{T}= & \sqrt{E_{V,p}^{2}-\mathrm{Re}\widetilde{\Sigma}_{\perp}(p_{\mathrm{on}})}-E_{p}^{V},\nonumber \\
\Delta E_{L}= & \sqrt{E_{V,p}^{2}-\frac{m_{V}^{2}}{|\mathbf{p}|^{2}}\mathrm{Re}\widetilde{\Sigma}_{00}(p_{\mathrm{on}})}-E_{p}^{V}.\label{eq:width-delta-e}
\end{align}
We see in Eq. (\ref{eq:Breit-Wigner}) that $\rho_{T/L}^{\mathrm{pole}}(p)$
denote pole contributions while $\rho_{T/L}^{\mathrm{cut}}(p)$ denote
cut contributions.

% done, QW, 2023.11.19, 8:00

\begin{figure}
\begin{centering}
\includegraphics[scale=0.6]{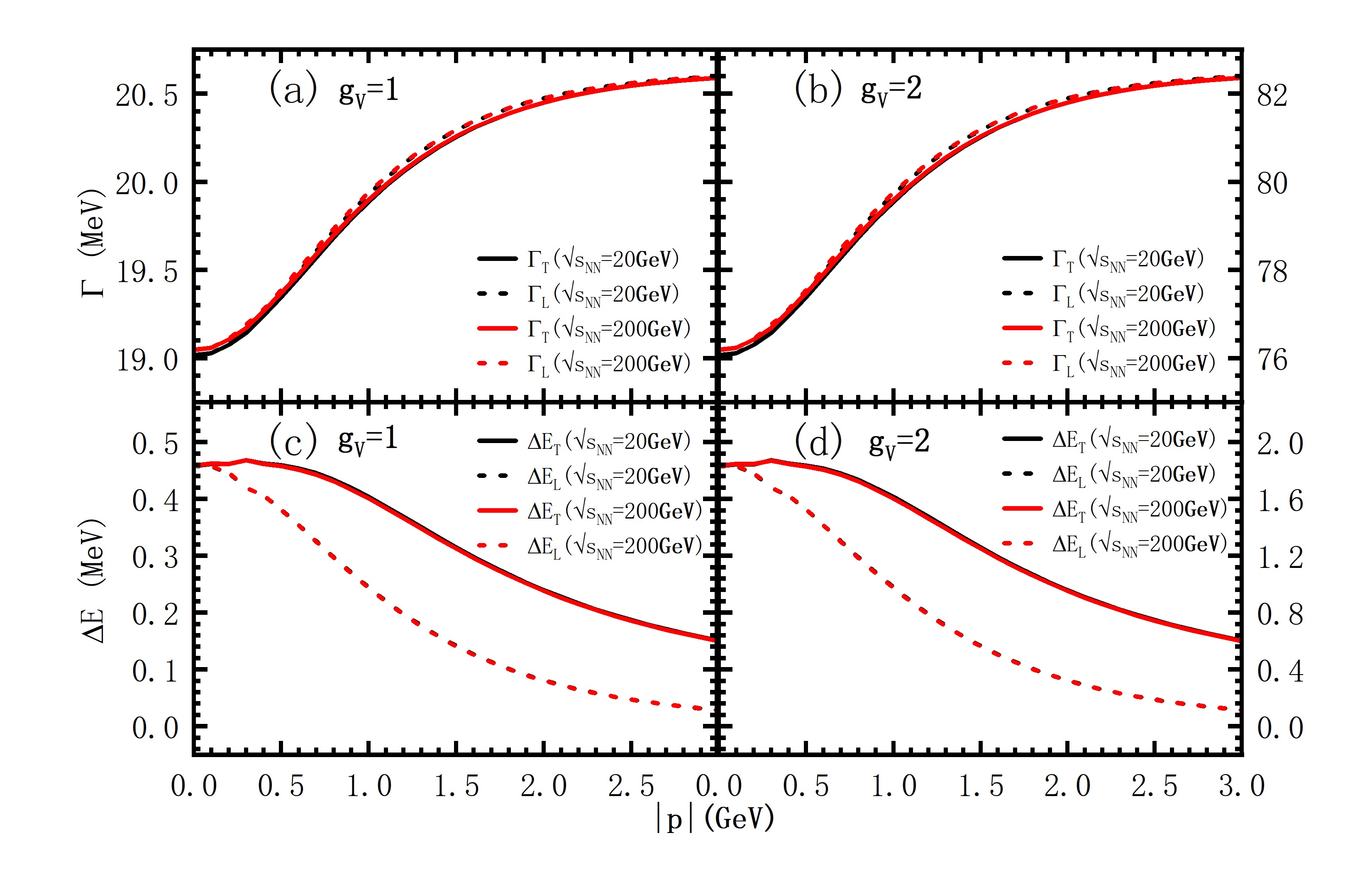}
\par\end{centering}
\caption{The width $\Gamma$ (a,b) and energy shift $\Delta E$ (c,d) for transverse
(solid lines) and longitudinal (dashed lines) modes as functions of
$|{\bf p}|$ at $g_{V}=1$ (a,c) and $g_{V}=2$ (b,d). Two sets of
values are chosen for the s-quark chemical potential and temperature
corresponding to the freezeout conditions at $\sqrt{s_{NN}}\approx20$
GeV and 200 GeV: $\mu_{s}=64.5$ MeV, $T=155.7$ MeV (black) and $\mu_{s}=7.4$
MeV and $T=158.4$ MeV (red). \label{fig:DeltaE-and-Gamma}}
\end{figure}

We plot widths and energy shifts in Fig. \ref{fig:DeltaE-and-Gamma}
as functions of $|{\bf p}|$ at $g_{V}=1,2$. We choose two sets of
values for the strange quark chemical potential and temperature corresponding
to the freezeout conditions at $\sqrt{s_{NN}}\approx$20 and 200 GeV
in heavy-ion collisions \citep{Andronic:2009gj,Andronic:2017pug}:
$\mu_{s}\approx\mu_{B}/3\approx64.5$ MeV and $T\approx155.7$ MeV
(black) and $\mu_{s}\approx\mu_{B}/3\approx7.4$ MeV and $T\approx158.4$
MeV (red). Other parameters are set to $g_{V}=1$, $m_{V}=1.02$ GeV,
and $m_{s}=419$ MeV. We can check $\rho_{T/L}^{\mathrm{pole}}(p)=0$
for these values of parameters since $4m_{q}^{2}-p^{2}<0$ at the
corrected mass-shell $p^{2}=m_{V}^{2}-\mathrm{Re}\widetilde{\Sigma}_{\perp}(p_{\mathrm{on}})$
and $p^{2}=m_{V}^{2}-(m_{V}^{2}/|\mathbf{p}|^{2})\mathrm{Re}\widetilde{\Sigma}_{00}(p_{\mathrm{on}})$
for transverse and longitudinal modes respectively. One can see in
Fig. \ref{fig:DeltaE-and-Gamma} that the width and energy shift are
almost independent of freezeout conditions at the collision energy
20 and 200 GeV.

% done, QW, 2023.11.24, 3:30

We find that the $\Gamma_{T/L}$ and $\Delta E_{T/L}$ are much smaller
than $m_{V}$, which allows us to introduce the following power counting
scheme 
\begin{equation}
\frac{\Delta E_{T/L}}{E_{p}^{V}}\sim\frac{\Gamma_{T/L}}{E_{p}^{V}}\sim\epsilon\ll1,
\end{equation}
where we have introduced $\epsilon$ as a small power counting parameter.
Since $\Gamma_{T}$ and $\Gamma_{L}$ are positive definite, we expect
that their difference is a second-order contribution $\Delta\Gamma/E_{p}^{V}\equiv(\Gamma_{T}-\Gamma_{L})/E_{p}^{V}\sim\mathcal{O}(\epsilon^{2})$,
while $E_{p}^{V}(1/\Gamma_{T}-1/\Gamma_{L})=E_{p}^{V}(\Gamma_{L}-\Gamma_{T})/(\Gamma_{T}\Gamma_{L})\sim\mathcal{O}(1)$.
On the other hand, such a cancellation may not happen for $\Delta E_{T}$
and $\Delta E_{L}$, because they may have different signs. Therefore
$(\Delta E_{T}-\Delta E_{L})/E_{p}^{V}\lesssim\mathcal{O}(\epsilon)$
could be a first-order contribution. According to hydrodynamic simulation
of the strong interaction matter in heavy-ion collisions, the thermal
shear tensor $\xi\equiv|\xi_{\gamma\lambda}|$ is a small quantity
of $\mathcal{O}(10^{-2})$, which can be treated as another power
counting parameter. With Eq. (\ref{eq:Breit-Wigner}) for spectral
functions, one can prove that the term with the $p_{0}$ integral
of $\delta G_{<}^{\mu\nu}(x,p)$ in the denominator of the right-hand-side
of Eq. (\ref{eq:dev-rho00}) is of the order $\xi E_{p}^{V}/\Gamma_{T/L}\sim\xi/\epsilon$,
while the term with the $p_{0}$ integral of $G_{<}^{\mu\nu}(x,p)$
is $\mathcal{O}(1)$. In order for the linear response theory to work,
one has to require $\xi/\epsilon\ll1$. 

% done, QW, 2023.11.24, 3:00

It is clear that the integrands in Eq. (\ref{eq:dev-rho00}) are suppressed
by spectral functions in the region of $p^{\mu}$ far from the mass-shell.
Therefore we can make an approximation by expanding $p_{0}$ in the
integrands around the on-shell energy $E_{p}^{V}$ in powers of $\delta p_{0}=p_{0}-E_{p}^{V}$
except spectral functions. To the first order in $\delta p_{0}$,
the $p_{0}$ integral of $G_{<}^{\mu\nu}(p)$ gives %
\begin{comment}
\begin{align}
\int_{0}^{+\infty}dp_{0}G_{<}^{\mu\nu}(p)= & \pi\sum_{a=T,L}\theta\left(4m_{q}^{2}+|\mathbf{p}|^{2}-(E_{p}^{V}+\Delta E_{a})^{2}\right)\frac{n_{B}(E_{p}^{V}+\Delta E_{a})}{E_{p}^{V}+\Delta E_{a}}\Delta_{a}^{\mu\nu}(p_{\text{on}})\nonumber \\
 & -2\sum_{a=T,L}\int_{0}^{+\infty}dp_{0}\rho_{a}^{\mathrm{cut}}(p)\left(1+\delta p_{0}\frac{\partial}{\partial E_{p}^{V}}\right)\left[n_{B}(p_{\text{on}})\Delta_{a}^{\mu\nu}(p_{\text{on}})\right],\label{eq:int_G<-1}
\end{align}
\end{comment}
\begin{equation}
\int_{0}^{+\infty}dp_{0}G_{<}^{\mu\nu}(p)=-2\sum_{a=T,L}\int_{0}^{+\infty}dp_{0}\rho_{a}^{\mathrm{cut}}(p)\left(1+\delta p_{0}\frac{\partial}{\partial E_{p}^{V}}\right)\left[n_{B}(p_{\text{on}})\Delta_{a}^{\mu\nu}(p_{\text{on}})\right],\label{eq:int_G<}
\end{equation}
while the $p_{0}$ integral of $\delta G_{<}^{\mu\nu}(p)$ from the
linear response to the shear tensor gives 
\begin{eqnarray}
\int_{0}^{+\infty}dp_{0}\delta G_{<}^{\mu\nu}(x,p) & = & 2T\xi_{\gamma\lambda}\sum_{a,b=L,T}\int_{0}^{+\infty}dp_{0}\rho_{a}^{\mathrm{cut}}(p)\rho_{b}^{\mathrm{cut}}(p)\nonumber \\
 &  & \times\left(1+\delta p_{0}\frac{\partial}{\partial E_{p}^{V}}\right)\left[\frac{\partial n_{B}(p_{\text{on}})}{\partial E_{p}^{V}}I_{ab}^{\mu\nu\gamma\lambda}(p_{\text{on}},p_{\text{on}})\right].\label{eq:int_delta_G}
\end{eqnarray}
Detailed calculations for the integrand in Eq. (\ref{eq:int_delta_G})
are given in Appendix \ref{sec:expansion-of-tensor}. The integrals
over $p_{0}$ in Eqs. (\ref{eq:int_G<}) and (\ref{eq:int_delta_G})
can be completed and the results are listed in Appendix \ref{sec:Integrals-for-spectral}.
Then $\delta\rho_{00}({\bf p})$ is calculated by substituting Eqs.
(\ref{eq:int_G<}) and (\ref{eq:int_delta_G}) into Eq. (\ref{eq:dev-rho00}).
Up to linear order in $\epsilon$ or $\xi$, the result reads 
\begin{eqnarray}
\delta\rho_{00}({\bf p}) & \approx & -\frac{1}{3}\left[1+n_{B}(E_{p}^{V})\right]\left\{ -L_{\mu\nu}(p_{\text{on}})\Delta_{T}^{\mu\nu}(p_{\text{on}})C_{0}({\bf p})\right.\nonumber \\
 &  & +\xi_{\gamma\lambda}L_{\mu\nu}(p_{\text{on}})\Delta_{T}^{\mu\nu}(p_{\text{on}})\nonumber \\
 &  & \times\left[\frac{p_{\text{on}}^{\gamma}p_{\text{on}}^{\lambda}}{(E_{p}^{V})^{2}}C_{1}({\bf p})+\frac{g^{\lambda0}p_{\text{on}}^{\gamma}+g^{\gamma0}p_{\text{on}}^{\lambda}-E_{p}^{V}g^{\gamma\lambda}}{2E_{p}^{V}}\left(C_{T}({\bf p})-C_{L}({\bf p})\right)\right]\nonumber \\
 &  & +\xi_{\gamma\lambda}L_{\mu\nu}(p_{\text{on}})\left[\Delta_{T}^{\gamma\nu}(p_{\text{on}})\Delta_{L}^{\lambda\mu}(p_{\text{on}})+\Delta_{L}^{\gamma\nu}(p_{\text{on}})\Delta_{T}^{\lambda\mu}(p_{\text{on}})\right]C_{2}({\bf p})\nonumber \\
 &  & \left.+\xi_{\gamma\lambda}L_{\mu\nu}(p_{\text{on}})\left[\Delta_{L}^{\gamma\nu}(p_{\text{on}})\Delta_{L}^{\lambda\mu}(p_{\text{on}})C_{L}({\bf p})+\Delta_{T}^{\gamma\nu}(p_{\text{on}})\Delta_{T}^{\lambda\mu}(p_{\text{on}})C_{T}({\bf p})\right]\right\} +\mathcal{O}(\epsilon^{2})\label{eq:deviation-rho00}
\end{eqnarray}
where the dimensionless coefficients are defined as 
\begin{eqnarray}
C_{0} & = & \frac{1+n_{B}(E_{p}^{V})+T/E_{p}^{V}}{1+n_{B}(E_{p}^{V})}\frac{\Delta E_{T}-\Delta E_{L}}{T},\nonumber \\
C_{1} & = & \frac{(E_{p}^{V})^{2}}{m_{V}}\left(\frac{1}{\Gamma_{T}}-\frac{1}{\Gamma_{L}}\right)+n_{B}(E_{p}^{V})\frac{(E_{p}^{V})^{2}}{m_{V}T}(\frac{\Delta E_{L}}{\Gamma_{L}}-\frac{\Delta E_{T}}{\Gamma_{T}}),\nonumber \\
C_{2} & = & \frac{4m_{V}E_{p}^{V}(\Gamma_{L}\Delta E_{T}+\Gamma_{T}\Delta E_{L})}{4(E_{p}^{V})^{2}(\Delta E_{T}-\Delta E_{L})^{2}+m_{V}^{2}(\Gamma_{L}+\Gamma_{T})^{2}},\nonumber \\
C_{T/L} & = & \frac{2E_{p}^{V}\Delta E_{T/L}}{m_{V}\Gamma_{T/L}}.\label{eq:dimless-coeff}
\end{eqnarray}
Noting that $\rho_{00}$ could deviate from $1/3$ due to a nonzero
$C_{0}$ independent of the shear tensor. Such a deviation arises
from the possible difference between spectral functions for transverse
and longitudinal modes \citep{Kim:2019ybi}. In the power counting
scheme, we can check that $C_{0}\sim\mathcal{O}(\epsilon)$ and other
coefficients $C_{i}$ with $i=1,2,T,L$ are all $\mathcal{O}(1)$.
The numerical results show that $C_{0}\sim\mathcal{O}(10^{-3})$ and
other coefficients $C_{i}$ with $i=1,2,T,L$ are $\mathcal{O}(10^{-1}\sim10^{-2})$
for $g_{V}=1,2$. The dominant term that is proportional to the shear
tensor is the $C_{1}$ term, which is controlled by $1/\Gamma_{T}-1/\Gamma_{L}$
for the current values of $g_{V}$.

% done, QW, 2023.11.22, 14:00

\begin{figure}
\includegraphics[scale=0.35]{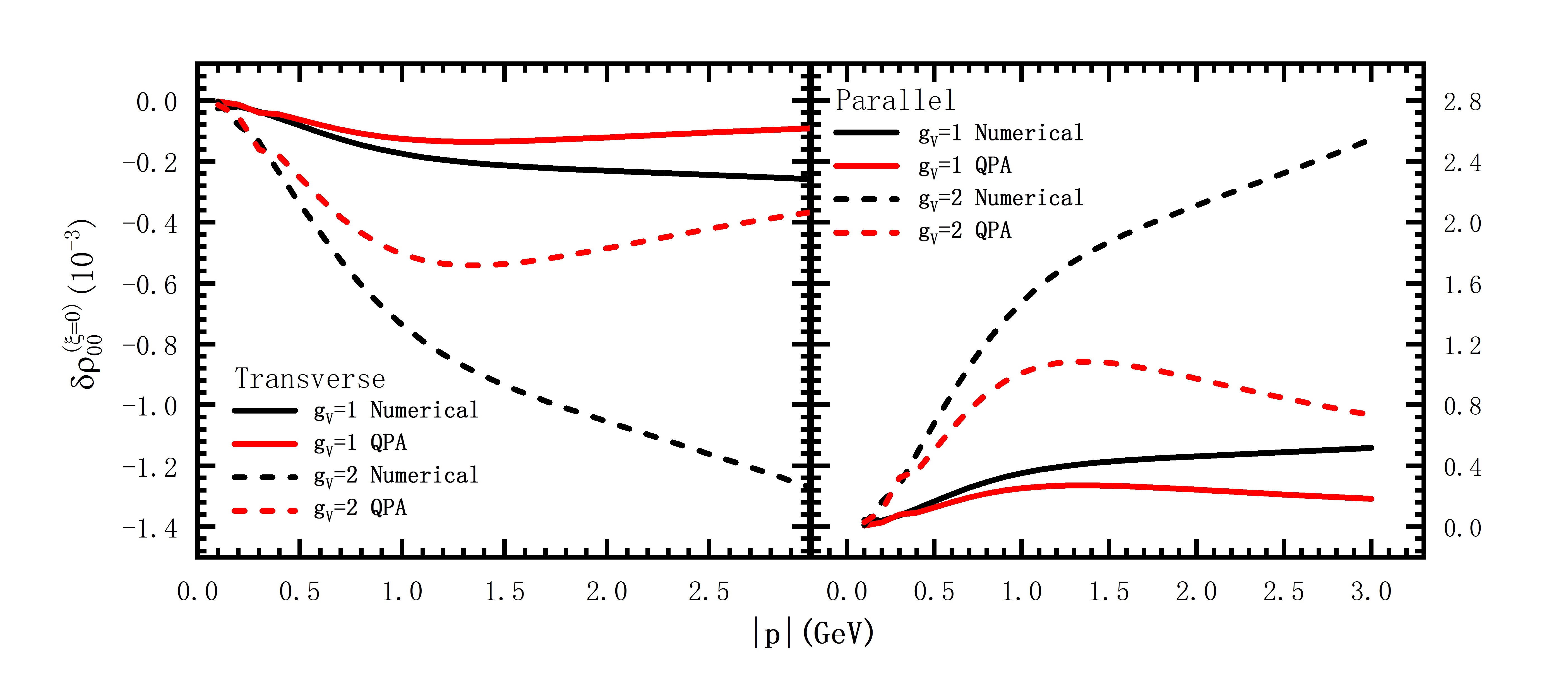}\caption{The numerical results for $\delta\rho_{00}^{(\xi=0)}$ in Eq. (\ref{eq:delta-rho-00})
for the transverse (left) and parallel (right) configurations in which
the momentum is transverse and parallel to the spin quantization direction
$z$ respectively. The results under the quasi-particle approximation
(QPA) using Eq. (\ref{eq:deviation-rho00}) are shown for comparison.
\label{fig:exact-numerical-xi0}}
\end{figure}

\begin{figure}
\includegraphics[scale=0.4]{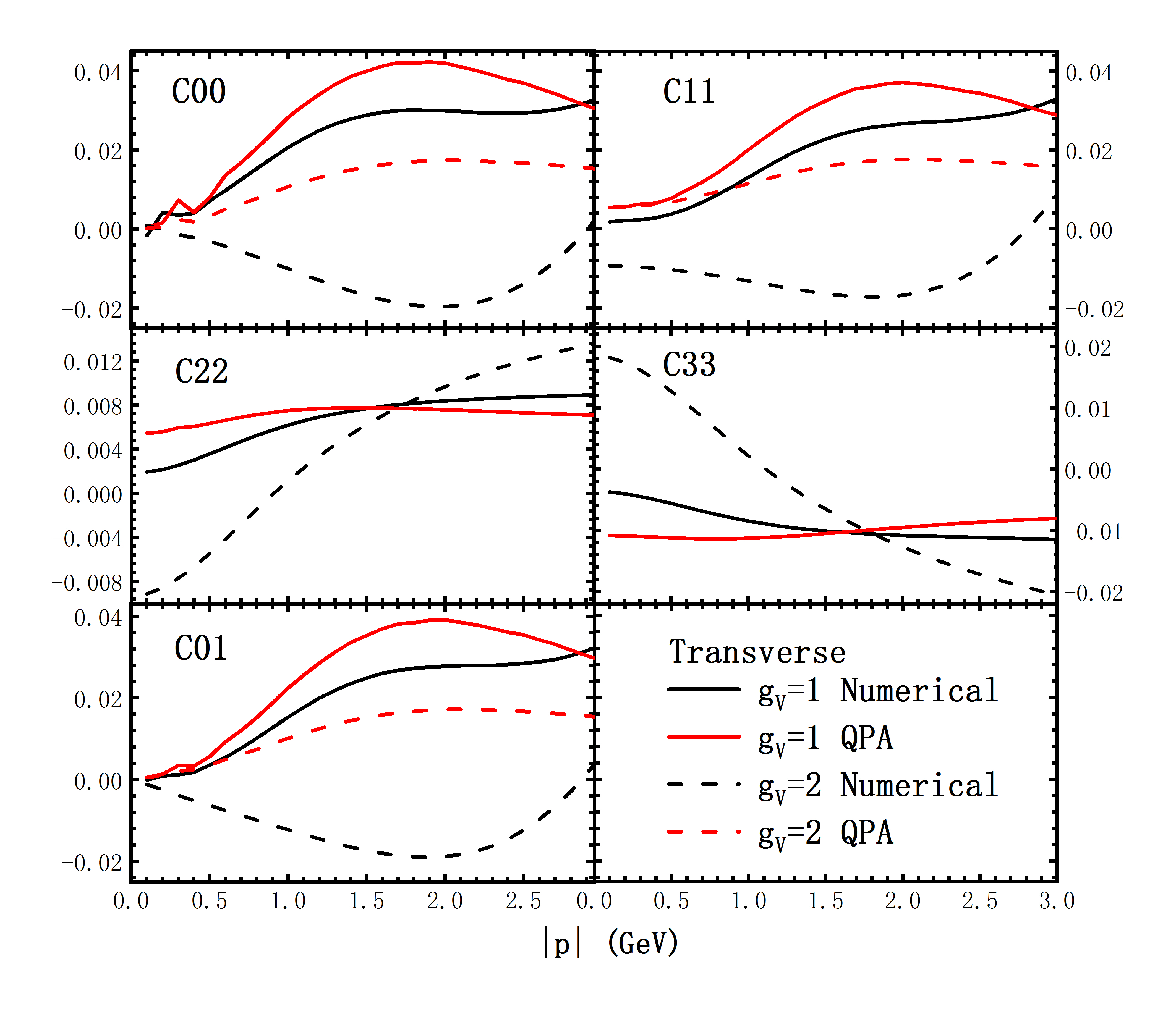}\caption{The numerical results for $C^{\mu\nu}$ in Eq. (\ref{eq:delta-rho-00})
for the transverse configuration in which the momentum is perpendicular
to the spin quantization direction $z$. The results under the quasi-particle
approximation (QPA) using Eq. (\ref{eq:deviation-rho00}) are shown
for comparison. \label{fig:exact-numerical-c-perp}}
\end{figure}

\begin{figure}
\includegraphics[scale=0.4]{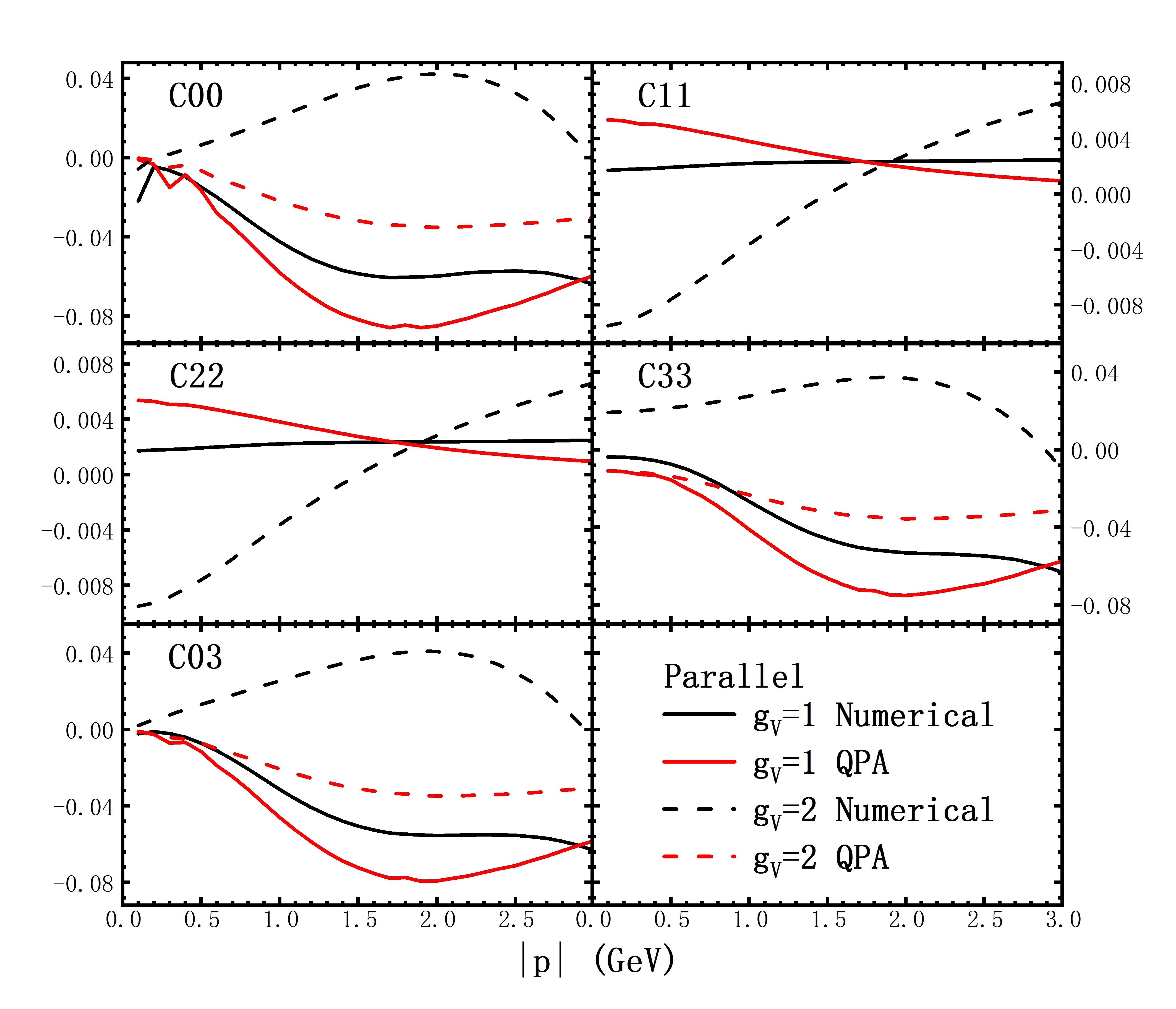}\caption{The numerical results for $C^{\mu\nu}$ in Eq. (\ref{eq:delta-rho-00})
for parallel directions of the spin quantization and momentum. The
spin quantization is chosen to be in the $z$ direction. The results
under the quasi-particle approximation (QPA) using Eq. (\ref{eq:deviation-rho00})
are shown for comparison. \label{fig:exact-numerical-c-para}}
\end{figure}

% done, QW, 2023.11.24, 3:30

\subsubsection{Numerical results}

In this subsection we will numerically calculate spectral functions
and $\delta\rho_{00}$ using Eqs. (\ref{eq:rho-tl}) and (\ref{eq:Im-self-energy-00-perp}).
We will compare numerical results with the QPA results using Eq. (\ref{eq:deviation-rho00}).
The parameters are set to the same values as in Subsection \ref{subsec:qpa}.
We can express $\delta\rho_{00}({\bf p})$ as 
\begin{equation}
\delta\rho_{00}({\bf p})=\delta\rho_{00}^{(\xi=0)}({\bf p})+\xi_{\mu\nu}C^{\mu\nu}({\bf p}),\label{eq:delta-rho-00}
\end{equation}
where $C^{\mu\nu}$ are dimensionless constants.

The numerical results for $\delta\rho_{00}^{(\xi=0)}({\bf p})$ are
shown in Fig. \ref{fig:exact-numerical-xi0}. The QPA results using
Eq. (\ref{eq:deviation-rho00}) are shown for comparison. We choose
two configurations for the mometum direction with respect to the spin
quantization one: transverse or parallel configuration. The analytic
results using Eq. (\ref{eq:deviation-rho00}) are also shown for comparison.
The results of the configuration with an arbitrary angle are between
these two limits. We see that the magnitude of $\delta\rho_{00}^{(\xi=0)}({\bf p})$
is about $10^{-3}$ for the values of parameters we choose.

The numerical results for the tensor coefficient $C^{\mu\nu}({\bf p})$
are shown in Figs. \ref{fig:exact-numerical-c-perp} and \ref{fig:exact-numerical-c-para}
for transverse and parallel configurations respectively. The QPA results
using Eq. (\ref{eq:deviation-rho00}) are shown for comparison. We
see that the magnitude of $C^{\mu\nu}({\bf p})$ is about $10^{-2}\sim10^{-3}$
for the values of parameters we choose, which is consistent with the
result of Ref. \citep{Wagner:2022gza} in the order of magnitude.

% done, QW, 2024.2.15, 11:30

\section{Discussion and conclusion}

\label{sec:Conclusion-and-Discussion}We study thermal medium effects
for the spin alignment of vector mesons from the meson-quark interaction
in the thermalized QGP, in which quarks, antiquarks and vector mesons
are assumed to be thermalized. Quarks and antiquarks are also assumed
to be unpolarized. We calculate the retarded self-energy of the vector
meson from the quark loop. The spectral function can be obtained from
the retarded two-point Green's function including the contribution
of the retarded self-energy. Other types of two-point Green's functions
with interaction can all be expressed in spectral functions. Then
we calculate the linear response of the two-point Green's function
to the thermal shear tensor using the Kubo formula, which provides
a correction to the Green's function. Such an effect is caused by
interaction.

% done, QW, 2023.11.22, 15:30

Finally the correction to $\rho_{00}$ can be expressed in terms of
spectral functions through one-loop self-energies. In order to obtain
an analytical formula for the correction to $\rho_{00}$, we take
the quasi-particle approximation: (a) the energy shifts and widths
from real and imaginary parts of self-energies are much smaller than
energies of vector mesons; (b) the difference between widths for transverse
and longitudinal modes is much smaller than widths themselves. This
approximation is supported by numerical results with the parameters
we have chosen. Under this approximation we derive an analytical formula
for the correction to $\rho_{00}$ to the linear order in the expansion
parameter in terms of energy shifts and widths. The numerical results
show that dimensionless coefficients of the thermal shear tensor are
presented as functions of scalar momentum with the magnitude of $\mathcal{O}(10^{-2}\sim10^{-3})$
for the chosen values of quark-meson coupling constant. The magnitude
of the contribution from the thermal shear tensor to $\rho_{00}$
is then $\mathcal{O}(10^{-4}\sim10^{-5})$ if the thermal shear tensor
is $\mathcal{O}(10^{-2})$.

Our results are based the one-loop self-energy with meson-quark interaction
in the QGP. One can also consider other interactions, such as $\rho\pi\pi$
or $\phi KK$ couplings, in the nuclear matter \citep{Kim:2019ybi,Kim:2022eku,Park:2022ayr}.

% done, QW, 2023.11.24, 3:40
\begin{acknowledgments}
The authors thank F. Li, S. Liu and S. Pu for helpful discussions.
This work is supported in part by the National Natural Science Foundation
of China (NSFC) under Grants No. 12135011 and by the Strategic Priority
Research Program of the Chinese Academy of Sciences (CAS) under Grant
No. XDB34030102.  
\end{acknowledgments}

\appendix

\section{Quark propagators}

\label{sec:quark-propagators}The propagators for unpolarized quarks
at the leading order are given by 
\begin{eqnarray}
S_{\alpha\beta}^{(0)<}(x,p) & = & (p\cdot\gamma+m_{q})_{\alpha\beta}\frac{\pi}{E_{p}}\nonumber \\
 &  & \times\left\{ -\delta(p_{0}-E_{p})f_{FD}^{(+)}(E_{p})+\delta(p_{0}+E_{p})\left[1-f_{FD}^{(-)}(E_{p})\right]\right\} ,\nonumber \\
S_{\alpha\beta}^{(0)>}(x,p) & = & (p\cdot\gamma+m_{q})_{\alpha\beta}\frac{\pi}{E_{p}}\nonumber \\
 &  & \times\left\{ \delta(p_{0}-E_{p})\left[1-f_{FD}^{(+)}(E_{p})\right]-\delta(p_{0}+E_{p})f_{FD}^{(-)}(E_{p})\right\} ,\nonumber \\
S_{\alpha\beta}^{(0)F}(x,p) & = & \frac{i(p\cdot\gamma+m_{q})_{\alpha\beta}}{p^{2}-m_{q}^{2}+i\epsilon}-\frac{\pi}{E_{p}}(p\cdot\gamma+m_{q})_{\alpha\beta}\nonumber \\
 &  & \times\left[\delta(p_{0}-E_{p})f_{FD}^{(+)}(E_{p})+\delta(p_{0}+E_{p})f_{FD}^{(-)}(E_{p})\right],\nonumber \\
S_{\alpha\beta}^{(0)A}(x,p) & = & \frac{i(p\cdot\gamma+m_{q})_{\alpha\beta}}{p^{2}-m_{q}^{2}-ip_{0}\epsilon},\nonumber \\
S_{\alpha\beta}^{(0)R}(x,p) & = & \frac{i(p\cdot\gamma+m_{q})_{\alpha\beta}}{p^{2}-m_{q}^{2}+ip_{0}\epsilon},\label{eq:quark-propogator}
\end{eqnarray}
where $\epsilon$ is a small positive number, $m_{q}=m_{\overline{q}}$
is the quark mass, and 
\begin{equation}
f_{FD}^{\pm}(E_{k})=\frac{1}{\exp\left(\beta E_{k}-\beta\mu_{q/\bar{q}}\right)+1},\label{eq:fermi-dirac-dist}
\end{equation}
are Fermi-Dirac distributions for quarks/antiquarks as functions of
the energy $E_{k}=\sqrt{|\mathbf{k}|^{2}+m_{q}^{2}}$ and chemical
potentials $\mu_{q/\bar{q}}$.

% done, QW, 2023.10.3, 11:00

\section{Retarded self-energy}

\label{sec:retarded-self-en}We evaluate the retarded self-energy
in Eq. (\ref{eq:self-energy-1}) using quark propagators in (\ref{eq:quark-propogator}).
The result reads 
\begin{eqnarray}
\Sigma_{R}^{\mu\nu}(p) & = & g_{V}^{2}\int\frac{d^{4}k}{(2\pi)^{4}}\left\{ \mathrm{Tr}\left[\gamma^{\mu}S_{F}(k)\gamma^{\nu}S_{F}(k-p)\right]-\mathrm{Tr}\left[\gamma^{\mu}S_{<}(k)\gamma^{\nu}S_{>}(k-p)\right]\right\} \nonumber \\
 & = & -g_{V}^{2}\int\frac{d^{4}k}{(2\pi)^{4}}\mathrm{Tr}\left\{ \gamma^{\mu}(k\cdot\gamma+m)\gamma^{\nu}\left[(k-p)\cdot\gamma+m\right]\right\} \nonumber \\
 &  & \times\frac{i}{(k-p)^{2}-m^{2}-i(k_{0}-p_{0})\epsilon}\nonumber \\
 &  & \times\frac{\pi}{E_{k}}\left\{ \delta(k_{0}-E_{k})f_{FD}^{(+)}(E_{k})+\delta(k_{0}+E_{k})f_{FD}^{(-)}(E_{k})\right\} \nonumber \\
 &  & +(\mu\leftrightarrow\nu,p\rightarrow-p,\epsilon\rightarrow-\epsilon)\nonumber \\
 &  & +g_{V}^{2}\int\frac{d^{4}k}{(2\pi)^{4}}\mathrm{Tr}\left\{ \gamma^{\mu}(k\cdot\gamma+m_{q})\gamma^{\nu}\left[(k-p)\cdot\gamma+m_{q}\right]\right\} \nonumber \\
 &  & \times\left[\frac{i}{k^{2}-m_{q}^{2}+i\epsilon}\cdot\frac{i}{(k-p)^{2}-m_{q}^{2}+i\epsilon}-\frac{\pi^{2}}{E_{k}E_{k-p}}\delta(k^{0}+E_{k})\delta(k^{0}-p^{0}-E_{k-p})\right]\nonumber \\
 & = & -ig_{V}^{2}\frac{1}{4\pi^{3}}(2I_{1}^{\mu\nu}+I_{2}^{\mu\nu})-ig_{V}^{2}I_{\mathrm{vac}}^{\mu\nu},\label{eq:retarded-self-en-i1i2}
\end{eqnarray}
where $I_{1}^{\mu\nu}$, $I_{2}^{\mu\nu}$ and $I_{\mathrm{vac}}^{\mu\nu}$
are defined as 
\begin{eqnarray}
I_{1}^{\mu\nu} & = & \int d^{3}kk_{\mathrm{on}}^{\mu}k_{\mathrm{on}}^{\nu}\frac{1}{E_{k}}\left[f_{FD}^{(+)}(E_{k})+f_{FD}^{(-)}(E_{k})\right]\nonumber \\
 &  & \times\left[\frac{1}{(k_{\mathrm{on}}+p)^{2}-m^{2}+i(E_{k}+p_{0})\epsilon}+\frac{1}{(k_{\mathrm{on}}-p)^{2}-m^{2}-i(E_{k}-p_{0})\epsilon}\right],\nonumber \\
I_{2}^{\mu\nu} & = & \int d^{3}k\frac{1}{E_{k}}\left[p^{\mu}k_{on}^{\nu}+p^{\nu}k_{\mathrm{on}}^{\mu}-g^{\mu\nu}(p\cdot k_{\mathrm{on}})\right]\left[f_{FD}^{(+)}(E_{k})+f_{FD}^{(-)}(E_{k})\right]\nonumber \\
 &  & \times\left[\frac{1}{(k_{\mathrm{on}}+p)^{2}-m^{2}+i(E_{k}+p_{0})\epsilon}-\frac{1}{(k_{\mathrm{on}}-p)^{2}-m^{2}-i(E_{k}-p_{0})\epsilon}\right],\\
I_{\mathrm{vac}}^{\mu\nu} & = & i\int\frac{d^{4}k}{(2\pi)^{4}}\mathrm{Tr}\left[\gamma^{\mu}(k\cdot\gamma+m_{q})\gamma^{\nu}((k-p)\cdot\gamma+m_{q})\right]\nonumber \\
 &  & \times\left[\frac{i}{k^{2}-m_{q}^{2}+i\epsilon}\cdot\frac{i}{(k-p)^{2}-m_{q}^{2}+i\epsilon}-\frac{\pi^{2}}{E_{k}E_{k-p}}\delta(k^{0}+E_{k})\delta(k^{0}-p^{0}-E_{k-p})\right],\label{eq:i1i2-munu}
\end{eqnarray}
where $k_{\mathrm{on}}^{\mu}=(E_{k},\mathbf{k})$ is an on-shell momentum
for the quark or anti-quark. The tensors $I_{1}^{\mu\nu}$ and $I_{2}^{\mu\nu}$
can be expressed in special functions $J_{\pm}(p;n_{1},n_{2})$ and
$J_{0}(n_{1},n_{2})$ defined as 
\begin{eqnarray}
J_{\pm}(p;n_{1},n_{2}) & \equiv & \int_{0}^{\infty}d|\mathbf{k}|\;E_{k}^{n_{1}}|\mathbf{k}|^{n_{2}}\left[f_{FD}^{(+)}(E_{k})+f_{FD}^{(-)}(E_{k})\right]\nonumber \\
 &  & \times\ln\frac{p^{2}\pm2p_{0}E_{k}+2|\mathbf{k}||\mathbf{p}|\pm i(E_{k}\pm p_{0})\epsilon}{p^{2}\pm2p_{0}E_{k}-2|\mathbf{k}||\mathbf{p}|\pm i(E_{k}\pm p_{0})\epsilon},\nonumber \\
J_{0}(n_{1},n_{2}) & \equiv & \int_{0}^{\infty}d|\mathbf{k}|\;E_{k}^{n_{1}}|\mathbf{k}|^{n_{2}}\left[f_{FD}^{(+)}(E_{k})+f_{FD}^{(-)}(E_{k})\right].\label{eq:jn1n2}
\end{eqnarray}

We now evaluate each element of $I_{1}^{\mu\nu}$ separately. The
result for $I_{1}^{00}$ is 
\begin{eqnarray}
I_{1}^{00} & = & 2\pi\int d|\mathbf{k}||\mathbf{k}|^{2}E_{k}\left[f_{FD}^{(+)}(E_{k})+f_{FD}^{(-)}(E_{k})\right]\int_{-1}^{1}d\cos\theta\nonumber \\
 &  & \times\left[\frac{1}{p^{2}+2p_{0}E_{k}-2|\mathbf{k}||\mathbf{p}|\cos\theta+i(E_{k}+p_{0})\epsilon}\right.\nonumber \\
 &  & \left.+\frac{1}{p^{2}-2p_{0}E_{k}+2|\mathbf{k}||\mathbf{p}|\cos\theta-i(E_{k}-p_{0})\epsilon}\right]\nonumber \\
 & = & \pi\frac{1}{|\mathbf{p}|}\left[J_{+}(p;1,1)+J_{-}(p;1,1)\right].\label{eq:retarded-00}
\end{eqnarray}
In evaluating $I^{0i}$, we decompose the vector $\mathbf{k}$ into
the component parallel and perpendicular to $\mathbf{p}$ as $\mathbf{k}=\hat{\mathbf{p}}(\mathbf{k}\cdot\mathbf{p})+\mathbf{k}_{T}$
with $\mathbf{k}_{T}\cdot\mathbf{p}=0$. The integral over the component
perpendicular to $\mathbf{p}$ vanishes. The result for $I_{1}^{0i}$
is 
\begin{eqnarray}
I_{1}^{0i}=I_{1}^{i0} & = & 2\pi\hat{\mathbf{p}}_{i}\int d|\mathbf{k}|d\theta\sin\theta\cos\theta|\mathbf{k}|^{3}\left[f_{FD}^{(+)}(E_{k})+f_{FD}^{(-)}(E_{k})\right]\nonumber \\
 &  & \times\left[\frac{1}{p^{2}+2p_{0}E_{k}-2|\mathbf{k}||\mathbf{p}|\cos\theta+i(E_{k}+p_{0})\epsilon}\right.\nonumber \\
 &  & \left.+\frac{1}{p^{2}-2p_{0}E_{k}+2|\mathbf{k}||\mathbf{p}|\cos\theta-i(E_{k}-p_{0})\epsilon}\right]\nonumber \\
 & = & \frac{\pi p^{2}}{2|\mathbf{p}|^{2}}\hat{\mathbf{p}}_{i}\left[J_{+}(p;0,1)-J_{-}(p;0,1)\right]\nonumber \\
 &  & +\frac{\pi p_{0}}{|\mathbf{p}|^{2}}\hat{\mathbf{p}}_{i}\left[J_{+}(p;1,1)+J_{-}(p;1,1)\right].\label{eq:retarded-0i}
\end{eqnarray}
To evaluate $I^{ij}$, we notice that it is symmetric in $i$ and
$j$, so we can decomposes it into components proportional to $\hat{\mathbf{p}}_{i}\hat{\mathbf{p}}_{j}$
and $\delta_{ij}-\hat{\mathbf{p}}_{i}\hat{\mathbf{p}}_{j}$ using
\begin{equation}
\mathbf{k}_{i}\mathbf{k}_{j}\rightarrow\hat{\mathbf{p}}_{i}\hat{\mathbf{p}}_{j}(\mathbf{k}\cdot\mathbf{p})+\frac{1}{2}\mathbf{k}_{T}^{2}(\delta_{ij}-\hat{\mathbf{p}}_{i}\hat{\mathbf{p}}_{j}).
\end{equation}
Then we obtain the result for $I^{ij}$ as

\begin{eqnarray}
I_{1}^{ij} & = & 2\pi\hat{\mathbf{p}}_{i}\hat{\mathbf{p}}_{j}\int d|\mathbf{k}|\;\frac{1}{E_{k}}|\mathbf{k}|^{4}\nonumber \\
 &  & \times\int d\theta\sin\theta\cos^{2}\theta\left[f_{FD}^{(+)}(E_{k})+f_{FD}^{(-)}(E_{k})\right]\nonumber \\
 &  & \times\left[\frac{1}{p^{2}+2p_{0}E_{k}-2\mathbf{k}\cdot\mathbf{p}+i(E_{k}+p_{0})\epsilon}\right.\nonumber \\
 &  & \left.+\frac{1}{p^{2}-2p_{0}E_{k}+2\mathbf{k}\cdot\mathbf{p}-i(E_{k}-p_{0})\epsilon}\right]\nonumber \\
 &  & +\pi(\delta_{ij}-\hat{\mathbf{p}}_{i}\hat{\mathbf{p}}_{j})\int d|\mathbf{k}|\;\frac{1}{E_{k}}|\mathbf{k}|^{4}\nonumber \\
 &  & \times\int d\theta\sin\theta\sin^{2}\theta\left[f_{FD}^{(+)}(E_{k})+f_{FD}^{(-)}(E_{k})\right]\nonumber \\
 &  & \times\left[\frac{1}{p^{2}+2p_{0}E_{k}-2\mathbf{k}\cdot\mathbf{p}+i(E_{k}+p_{0})\epsilon}\right.\nonumber \\
 &  & \left.+\frac{1}{p^{2}-2p_{0}E_{k}+2\mathbf{k}\cdot\mathbf{p}-i(E_{k}-p_{0})\epsilon}\right]\nonumber \\
 & = & \hat{\mathbf{p}}_{i}\hat{\mathbf{p}}_{j}\frac{\pi}{4|\mathbf{p}|^{3}}\left\{ -8p^{2}|\mathbf{p}|J_{0}(-1,2)+p^{4}\left[J_{+}(p;-1,1)+J_{-}(p;-1,1)\right]\right.\nonumber \\
 &  & \left.+4p_{0}^{2}\left[J_{+}(p;1,1)+J_{-}(p;1,1)\right]+4p_{0}p^{2}\left[J_{+}(p;0,1)-J_{-}(p;0,1)\right]\right\} \nonumber \\
 &  & +(\delta_{ij}-\hat{\mathbf{p}}_{i}\hat{\mathbf{p}}_{j})\frac{\pi}{8|\mathbf{p}|^{3}}\left\{ 8p^{2}|\mathbf{p}|J_{0}(-1,2)-p^{4}\left[J_{+}(p;-1,1)+J_{-}(p;-1,1)\right]\right.\nonumber \\
 &  & -4p_{0}^{2}\left[J_{+}(p;1,1)+J_{-}(p;1,1)\right]-4p_{0}p^{2}\left[J_{+}(p;0,1)-J_{-}(p;0,1)\right]\nonumber \\
 &  & \left.+4|\mathbf{p}|^{2}\left[J_{+}(p;-1,3)+J_{-}(p;-1,3)\right]\right\} .\label{eq:retarded-ij}
\end{eqnarray}
In Eqs. (\ref{eq:retarded-00}), (\ref{eq:retarded-0i}) and (\ref{eq:retarded-ij})
we have express the result of $I_{1}^{\mu\nu}$ in terms of special
functions $J_{\pm}(p;n_{1},n_{2})$ and $J_{0}(n_{1},n_{2})$ in Eq.
(\ref{eq:jn1n2}).

The derivation of $I_{2}^{\mu\nu}$ is similar and straightforward.
Here we just list the result as follows 
\begin{eqnarray}
I_{2}^{00} & = & -4\pi J_{0}(-1,2)+2\pi\frac{p_{0}}{|\mathbf{p}|}\left[J_{+}(p;0,1)-J_{-}(p;0,1)\right]\nonumber \\
 &  & +\frac{\pi p^{2}}{2|\mathbf{p}|}\left[J_{+}(p;-1,1)+J_{-}(p;-1,1)\right],\nonumber \\
I_{2}^{0i} & = & I_{2}^{i0}=-4\pi\hat{\mathbf{p}}_{i}\frac{p_{0}}{|\mathbf{p}|}J_{0}(-1,2)+\frac{\pi}{2}\hat{\mathbf{p}}_{i}\frac{p_{0}p^{2}}{|\mathbf{p}|^{2}}\left[J_{+}(p;-1,1)+J_{-}(p;-1,1)\right]\nonumber \\
 &  & +\pi(1+\frac{p_{0}^{2}}{|\mathbf{p}|^{2}})\hat{\mathbf{p}}_{i}\left[J_{+}(p;0,1)-J_{-}(p;0,1)\right],\nonumber \\
I_{2}^{ij} & = & \hat{\mathbf{p}}_{i}\hat{\mathbf{p}}_{j}\left\{ -4\pi J_{0}(-1,2)+\frac{\pi}{2}\frac{p^{2}}{|\mathbf{p}|}\left[J_{+}(p;-1,1)+J_{-}(p;-1,1)\right]\right.\nonumber \\
 &  & \left.+2\pi\frac{p_{0}}{|\mathbf{p}|}\left[J_{+}(p;0,1)-J_{-}(p;0,1)\right]\right\} \nonumber \\
 &  & +(\delta_{ij}-\hat{\mathbf{p}}_{i}\hat{\mathbf{p}}_{j})\left\{ 4\pi J_{0}(-1,2)-\frac{\pi}{2}\frac{p^{2}}{|\mathbf{p}|}\left[J_{+}(p;-1,1)+J_{-}(p;-1,1)\right]\right\} .
\end{eqnarray}

Using the results for elements of $I_{1}^{\mu\nu}$ and $I_{2}^{\mu\nu}$,
we obtain the elements of $2I_{1}^{\mu\nu}+I_{2}^{\mu\nu}$ in Eq.
(\ref{eq:retarded-self-en-i1i2}), 
\begin{eqnarray}
2I_{1}^{00}+I_{2}^{00} & = & -4\pi J_{0}(-1,2)+2\pi\frac{1}{|\mathbf{p}|}\left[J_{+}(p;1,1)+J_{-}(p;1,1)\right]\nonumber \\
 &  & +2\pi\frac{p_{0}}{|\mathbf{p}|}\left[J_{+}(p;0,1)-J_{-}(p;0,1)\right]\nonumber \\
 &  & +\frac{\pi p^{2}}{2|\mathbf{p}|}\left[J_{+}(p;-1,1)+J_{-}(p;-1,1)\right],\nonumber \\
2I_{1}^{0i}+I_{2}^{0i} & = & \hat{\mathbf{p}}_{i}\frac{p_{0}}{|\mathbf{p}|}(2I_{1}^{00}+I_{2}^{00})\nonumber \\
2I_{1}^{ij}+I_{2}^{ij} & = & \hat{\mathbf{p}}_{i}\hat{\mathbf{p}}_{j}\frac{p_{0}^{2}}{|\mathbf{p}|^{2}}(2I_{1}^{00}+I_{2}^{00})\nonumber \\
 &  & +(\delta_{ij}-\hat{\mathbf{p}}_{i}\hat{\mathbf{p}}_{j})\frac{\pi}{4|\mathbf{p}|^{3}}\left\{ 8(p_{0}^{2}+|\mathbf{p}|^{2})|\mathbf{p}|J_{0}(-1,2)\right.\nonumber \\
 &  & -p^{2}(p_{0}^{2}+|\mathbf{p}|^{2})\left[J_{+}(p;-1,1)+J_{-}(p;-1,1)\right]\nonumber \\
 &  & -4p_{0}^{2}\left[J_{+}(p;1,1)+J_{-}(p;1,1)\right]\nonumber \\
 &  & -4p_{0}p^{2}\left[J_{+}(p;0,1)-J_{-}(p;0,1)\right]\nonumber \\
 &  & \left.+4|\mathbf{p}|^{2}\left[J_{+}(p;-1,3)+J_{-}(p;-1,3)\right]\right\} .\label{eq:i1i2-rel}
\end{eqnarray}

The vacuum contribution $I_{\mathrm{vac}}^{\mu\nu}$ for $p^{0}>0$
can be evaluated by dimensional regularization as 
\begin{align}
I_{\mathrm{vac}}^{\mu\nu}= & i\int\frac{d^{4}k}{(2\pi)^{4}}\mathrm{Tr}\left\{ \gamma^{\mu}(k\cdot\gamma+m_{q})\gamma^{\nu}\left[(k-p)\cdot\gamma+m_{q}\right]\right\} \nonumber \\
 & \times\frac{i}{k^{2}-m_{q}^{2}+i\epsilon}\cdot\frac{i}{(k-p)^{2}-m_{q}^{2}+i\epsilon}\nonumber \\
= & (g^{\mu\nu}p^{2}-p^{\mu}p^{\nu})\frac{1}{2\pi^{2}}\int_{0}^{1}dxx(1-x)\nonumber \\
 & \times\left[\frac{2}{\epsilon}-\log(-x(1-x)p^{2}+m_{q}^{2}-i0^{+})-\gamma+\log(4\pi)+\mathcal{O}(\epsilon)\right]\nonumber \\
= & p^{2}\Delta^{\mu\nu}\frac{1}{2\pi^{2}}\int_{0}^{1}dxx(1-x)\nonumber \\
 & \times\left[\frac{2}{\epsilon}-\log\left|-x(1-x)p^{2}+m_{q}^{2}\right|-\gamma+\log(4\pi)+\mathcal{O}(\epsilon)\right]\nonumber \\
 & +ip^{2}\Delta^{\mu\nu}\frac{1}{2\pi^{2}}\int_{0}^{1}dxx(1-x)\pi\theta\left[x(1-x)p^{2}-m_{q}^{2}\right],
\end{align}
where $\epsilon=4-d$ ($d$ is an arbitrary space-time dimension in
regularization) and $\gamma\approx0.5772$. Here the term proportional
to delta functions in $I_{\mathrm{vac}}^{\mu\nu}$ is vanishing. The
real part of $I_{\mathrm{vac}}^{\mu\nu}$ can be canceled by introducing
a renormalization term with the condition $\mathrm{Re}I_{\mathrm{vac}}^{\mu\nu}(p^{2}=m_{V}^{2})=0$.
The imaginary part $I_{\mathrm{vac}}^{\mu\nu}$ is nonzero when $\sqrt{p^{2}}>2m_{q}$
and contributes to the spectral density. This corresponds to pair
production or annihilation processes. 

% done, QW, 2023.10.18, 10:00

\section{Expansion of part of integrand in Eq. (\ref{eq:G_<(1)_p0_integral})
at mass-shell}

\label{sec:expansion-of-tensor}In this appendix, we will expand $p_{0}$
around $E_{p}^{V}$ in powers of $\delta p_{0}$ for the integrand
in the second term of Eq. (\ref{eq:G_<(1)_p0_integral}) except spectral
functions. The integrand can be written as 
\begin{eqnarray}
I^{\mu\nu\gamma\lambda} & = & \frac{\partial n_{B}(p_{0})}{\partial p_{0}}\rho_{a}(p)\rho_{b}(p)I_{ab}^{\mu\nu\gamma\lambda}(p,p)\nonumber \\
 & = & \rho_{a}(p)\rho_{b}(p)\left(1+\delta p_{0}\frac{\partial}{\partial E_{p}^{V}}\right)\left[\frac{\partial n_{B}(p_{\mathrm{on}})}{\partial p_{\mathrm{on}}}I_{ab}^{\mu\nu\gamma\lambda}(p_{\mathrm{on}},p_{\mathrm{on}})\right]
\end{eqnarray}
where a summation over $a,b=L,T$ is implied.

First we expand $I_{ab}^{\mu\nu\gamma\lambda}(p,p)$ in (\ref{eq:I_ab})
at $p_{0}=E_{p}^{V}$. We can express $p^{\mu}=p_{\mathrm{on}}^{\mu}+\delta p_{0}g^{\mu0}$,
where $\delta p_{0}=p_{0}-E_{p}^{V}$. We note that $\Delta_{T}^{\mu\nu}(p)$
does not depend on $p_{0}$, but $\Delta_{L}^{\mu\nu}(p)=\Delta^{\mu\nu}(p)-\Delta_{T}^{\mu\nu}(p)$
depends on $p_{0}$ through $\Delta^{\mu\nu}$. Then to the first
order in $\delta p_{0}$ they can be expanded as 
\begin{align}
\Delta_{T}^{\mu\nu}(p)= & \Delta_{T}^{\mu\nu}(p_{\mathrm{on}}),\nonumber \\
\Delta^{\mu\nu}(p)\approx & \Delta^{\mu\nu}(p_{\mathrm{on}})+\frac{2E_{p}^{V}}{m_{V}^{4}}p_{\mathrm{on}}^{\mu}p_{\mathrm{on}}^{\nu}\delta p_{0}-\frac{1}{m_{V}^{2}}(p_{\mathrm{on}}^{\mu}g^{\nu0}+p_{\mathrm{on}}^{\nu}g^{\mu0})\delta p_{0},\nonumber \\
\Delta_{L}^{\mu\nu}(p)\approx & \Delta_{L}^{\mu\nu}(p_{\mathrm{on}})+\frac{2E_{p}^{V}}{m_{V}^{4}}p_{\mathrm{on}}^{\mu}p_{\mathrm{on}}^{\nu}\delta p_{0}-\frac{1}{m_{V}^{2}}(p_{\mathrm{on}}^{\mu}g^{\nu0}+p_{\mathrm{on}}^{\nu}g^{\mu0})\delta p_{0}.
\end{align}
Then $I_{ab}^{\mu\nu\gamma\lambda}(p,p)$ can be expanded to the leading
order in $\delta p_{0}$ as
\begin{align}
I_{ab}^{\mu\nu\gamma\lambda}(p,p)= & 2p^{\lambda}p^{\gamma}\Delta_{a,\alpha}^{\nu}(p)\Delta_{b}^{\mu\alpha}(p)\nonumber \\
 & +(p^{2}-m_{V}^{2})\left[\Delta_{a}^{\gamma\nu}(p)\Delta_{b}^{\mu\lambda}(p)+\Delta_{a}^{\lambda\nu}(p)\Delta_{b}^{\mu\gamma}(p)\right]\nonumber \\
 & -g^{\gamma\lambda}g_{\beta\alpha}(p^{2}-m_{V}^{2})\Delta_{a}^{\alpha\nu}(p)\Delta_{b}^{\mu\beta}(p),\nonumber \\
I_{ab}^{\mu\nu\gamma\lambda}(p_{\mathrm{on}},p_{\mathrm{on}})= & 2p_{\mathrm{on}}^{\lambda}p_{\mathrm{on}}^{\gamma}\Delta_{a,\alpha}^{\nu}(p_{\mathrm{on}})\Delta_{b}^{\mu\alpha}(p_{\mathrm{on}}),\nonumber \\
I_{TT}^{\mu\nu\gamma\lambda}(p,p)\approx & 2p_{\mathrm{on}}^{\lambda}p_{\mathrm{on}}^{\gamma}\Delta_{T}^{\mu\nu}(p)\nonumber \\
 & +2(p_{\mathrm{on}}^{\lambda}g^{\gamma0}+p_{\mathrm{on}}^{\gamma}g^{\lambda0}-E_{p}^{V}g^{\gamma\lambda})\delta p_{0}\Delta_{T}^{\mu\nu}(p)\nonumber \\
 & +2E_{p}^{V}\delta p_{0}\left[\Delta_{T}^{\gamma\nu}(p)\Delta_{T}^{\mu\lambda}(p)+\Delta_{T}^{\lambda\nu}(p)\Delta_{T}^{\mu\gamma}(p)\right],\nonumber \\
I_{TL}^{\mu\nu\gamma\lambda}(p,p)\approx & 2E_{p}^{V}\delta p_{0}\left[\Delta_{T}^{\gamma\nu}(p)\Delta_{L}^{\mu\lambda}(p_{\mathrm{on}})+\Delta_{T}^{\lambda\nu}(p)\Delta_{L}^{\mu\gamma}(p_{\mathrm{on}})\right],\nonumber \\
I_{LT}^{\mu\nu\gamma\lambda}(p,p)\approx & 2E_{p}^{V}\delta p_{0}\left[\Delta_{L}^{\gamma\nu}(p_{\mathrm{on}})\Delta_{T}^{\mu\lambda}(p)+\Delta_{L}^{\lambda\nu}(p_{\mathrm{on}})\Delta_{T}^{\mu\gamma}(p)\right],\nonumber \\
I_{LL}^{\mu\nu\gamma\lambda}(p,p)= & 2p_{\mathrm{on}}^{\lambda}p_{\mathrm{on}}^{\gamma}\Delta_{L}^{\mu\nu}(p_{\mathrm{on}})+4\frac{E_{p}^{V}}{m_{V}^{4}}p_{\mathrm{on}}^{\lambda}p_{\mathrm{on}}^{\gamma}p_{\mathrm{on}}^{\mu}p_{\mathrm{on}}^{\nu}\delta p_{0}\nonumber \\
 & -2\frac{1}{m_{V}^{2}}p_{\mathrm{on}}^{\lambda}p_{\mathrm{on}}^{\gamma}(p_{\mathrm{on}}^{\mu}g^{\nu0}+p_{\mathrm{on}}^{\nu}g^{\mu0})\delta p_{0}\nonumber \\
 & +2(p_{\mathrm{on}}^{\lambda}g^{\gamma0}+p_{\mathrm{on}}^{\gamma}g^{\lambda0}-E_{p}^{V}\eta^{\gamma\lambda})\delta p_{0}\Delta_{L}^{\mu\nu}(p_{\mathrm{on}})\nonumber \\
 & +2E_{p}^{V}\delta p_{0}\left[\Delta_{L}^{\gamma\nu}(p_{\mathrm{on}})\Delta_{L}^{\mu\lambda}(p_{\mathrm{on}})+\Delta_{L}^{\lambda\nu}(p_{\mathrm{on}})\Delta_{L}^{\mu\gamma}(p_{\mathrm{on}})\right].
\end{align}

The function $\partial n_{B}(p_{0})/\partial p_{0}$ is expanded to
the first order in $\delta p_{0}$ as 
\begin{align}
\frac{\partial n_{B}(p_{0})}{\partial p_{0}}\approx & \frac{\partial n_{B}(E_{p}^{V})}{\partial E_{p}^{V}}+\frac{\partial^{2}n_{B}(E_{p}^{V})}{\partial^{2}E_{p}^{V}}\delta p_{0}+\mathcal{O}[(\delta p_{0})^{2}]\nonumber \\
= & -\beta n_{B}(E_{p}^{V})\left[1+n_{B}(E_{p}^{V})\right]\nonumber \\
 & +\beta^{2}n_{B}(E_{p}^{V})\left[1+n_{B}(E_{p}^{V})\right]\left[1+2n_{B}(E_{p}^{V})\right]\delta p_{0},
\end{align}
To the first order in $\delta p_{0}$, the integrand is expanded as
\begin{eqnarray}
I_{\mathrm{LO}}^{\mu\nu\gamma\lambda} & = & \rho_{a}(p)\rho_{b}(p)\frac{\partial n_{B}(E_{p}^{V})}{\partial E_{p}^{V}}I_{ab}^{\mu\nu\gamma\lambda}(p_{\mathrm{on}},p_{\mathrm{on}})\nonumber \\
 & = & 2p_{\mathrm{on}}^{\lambda}p_{\mathrm{on}}^{\gamma}\frac{\partial n_{B}(E_{p}^{V})}{\partial E_{p}^{V}}\left\{ \Delta^{\mu\nu}(p_{\mathrm{on}})\rho_{L}^{2}(p)+\Delta_{T}^{\mu\nu}(p_{\mathrm{on}})\left[\rho_{T}^{2}(p)-\rho_{L}^{2}(p)\right]\right\} ,\label{eq:lo-tensor}\\
I_{\mathrm{NLO}}^{\mu\nu\gamma\lambda} & = & 2p_{\mathrm{on}}^{\lambda}p_{\mathrm{on}}^{\gamma}\frac{\partial^{2}n_{B}(E_{p}^{V})}{\partial^{2}E_{p}^{V}}\delta p_{0}\nonumber \\
 &  & \times\left\{ \Delta^{\mu\nu}(p_{\mathrm{on}})\rho_{L}^{2}(p)+\Delta_{T}^{\mu\nu}(p_{\mathrm{on}})\left[\rho_{T}^{2}(p)-\rho_{L}^{2}(p)\right]\right\} \nonumber \\
 &  & +2\delta p_{0}\rho_{T}^{2}(p)\frac{\partial n_{B}(E_{p}^{V})}{\partial E_{p}^{V}}\left\{ (p_{\mathrm{on}}^{\lambda}g^{\gamma0}+p_{\mathrm{on}}^{\gamma}g^{\lambda0}-E_{p}^{V}g^{\gamma\lambda})\Delta_{T}^{\mu\nu}(p)\right.\nonumber \\
 &  & \left.+E_{p}^{V}\left[\Delta_{T}^{\gamma\nu}(p)\Delta_{T}^{\mu\lambda}(p)+\Delta_{T}^{\lambda\nu}(p)\Delta_{T}^{\mu\gamma}(p)\right]\right\} \nonumber \\
 &  & +2\delta p_{0}\rho_{L}^{2}(p)\frac{\partial n_{B}(E_{p}^{V})}{\partial E_{p}^{V}}\left\{ 2\frac{E_{p}^{V}}{m_{V}^{4}}p_{\mathrm{on}}^{\lambda}p_{\mathrm{on}}^{\gamma}p_{\mathrm{on}}^{\mu}p_{\mathrm{on}}^{\nu}\right.\nonumber \\
 &  & -\frac{1}{m_{V}^{2}}p_{\mathrm{on}}^{\lambda}p_{\mathrm{on}}^{\gamma}(p_{\mathrm{on}}^{\mu}g^{\nu0}+p_{\mathrm{on}}^{\nu}g^{\mu0})+\Delta_{L}^{\mu\nu}(p_{\mathrm{on}})(p_{\mathrm{on}}^{\lambda}g^{\gamma0}+p_{\mathrm{on}}^{\gamma}g^{\lambda0}-E_{p}^{V}g^{\gamma\lambda})\nonumber \\
 &  & \left.+E_{p}^{V}\left[\Delta_{L}^{\gamma\nu}(p_{\mathrm{on}})\Delta_{L}^{\mu\lambda}(p_{\mathrm{on}})+\Delta_{L}^{\lambda\nu}(p_{\mathrm{on}})\Delta_{L}^{\mu\gamma}(p_{\mathrm{on}})\right]\right\} \nonumber \\
 &  & +2E_{p}^{V}\delta p_{0}\rho_{T}(p)\rho_{L}(p)\frac{\partial n_{B}(E_{p}^{V})}{\partial E_{p}^{V}}\left[\Delta_{T}^{\gamma\nu}(p)\Delta_{L}^{\mu\lambda}(p_{\mathrm{on}})+\Delta_{T}^{\lambda\nu}(p)\Delta_{L}^{\mu\gamma}(p_{\mathrm{on}})\right.\nonumber \\
 &  & \left.+\Delta_{T}^{\mu\lambda}(p)\Delta_{L}^{\gamma\nu}(p_{\mathrm{on}})+\Delta_{T}^{\mu\gamma}(p)\Delta_{L}^{\lambda\nu}(p_{\mathrm{on}})\right].\label{eq:nlo-tensor}
\end{eqnarray}

\section{Integrals for spectral functions}

\label{sec:Integrals-for-spectral}The integrals in Eqs. (\ref{eq:int_G<})
and ($\ref{eq:int_delta_G}$) are given as follows

\begin{eqnarray}
\int_{0}^{\infty}dp_{0}\rho_{L/T}(p) & \approx & \frac{\pi}{2E_{p}^{V}}\left(1-\frac{\Delta E_{L/T}}{E_{p}^{V}}\right)+\mathcal{O}(\epsilon^{2})\nonumber \\
\int_{0}^{\infty}dp_{0}\delta p_{0}\rho_{L/T}(p) & \approx & \frac{\pi}{2}\frac{\Delta E_{L/T}}{E_{p}^{V}}+\mathcal{O}(\epsilon^{2})\nonumber \\
\int_{0}^{\infty}dp_{0}\rho_{L/T}^{2}(p) & \approx & \frac{\pi}{4E_{p}m_{V}\Gamma_{L/T}}\left(1-\frac{\Delta E_{L/T}}{E_{p}}\right)+\mathcal{O}(\epsilon)\nonumber \\
\int_{0}^{\infty}dp_{0}\delta p_{0}\rho_{L/T}^{2}(p) & \approx & \frac{\pi\Delta E_{L/T}}{4E_{p}m_{V}\Gamma_{L/T}}+\mathcal{O}(\epsilon)\nonumber \\
\int_{0}^{\infty}dp_{0}\delta p_{0}\rho_{L}(p)\rho_{T}(p) & \approx & \frac{\pi}{2E_{p}}\frac{m_{V}(\Gamma_{L}\Delta E_{T}+\Gamma_{T}\Delta E_{L})}{\left[4E_{p}^{2}(\Delta E_{L}-\Delta E_{T})^{2}+m_{V}^{2}(\Gamma_{T}+\Gamma_{L})^{2}\right]}+\mathcal{O}(\epsilon)
\end{eqnarray}

\bibliographystyle{h-physrev}
\phantomsection\addcontentsline{toc}{section}{\refname}\bibliography{shear_induced_spin_alignment}

\end{document}